\documentclass[11pt,a4paper]{article}

\usepackage[T1]{fontenc}
\usepackage{lmodern}
\usepackage{microtype}
\usepackage{amsmath,amssymb,amsthm}
\usepackage{graphicx}
\usepackage{algorithm}
\usepackage{algpseudocode}
\usepackage{booktabs}
\usepackage{multirow}
\usepackage{xcolor}
\usepackage[font=small,labelfont=bf,labelsep=period,figurename=Fig.,tablename=Table]{caption}
\usepackage[margin=2.5cm,headsep=0.8cm]{geometry}

\usepackage{titlesec}
\titleformat{\section}{\large\bfseries}{\thesection}{0.8em}{}
\titleformat{\subsection}{\normalsize\bfseries}{\thesubsection}{0.8em}{}
\titleformat{\subsubsection}[runin]{\normalsize\bfseries}{\thesubsubsection}{0.5em}{}[.]

\usepackage{fancyhdr}
\fancypagestyle{first}{\fancyhf{}\fancyfoot[C]{\small\thepage}}

\usepackage[colorlinks,linkcolor=black,citecolor=blue!50!black,urlcolor=blue!50!black]{hyperref}

\theoremstyle{plain}
\newtheorem{thm}{Theorem}
\newtheorem{lemma}{Lemma}
\newtheorem{proposition}{Proposition}
\newtheorem{corollary}{Corollary}
\theoremstyle{definition}
\newtheorem{assumption}{Assumption}
\newtheorem{definition}{Definition}
\newtheorem{remark}{Remark}

\newenvironment{pf}{\par\addvspace{0.5\baselineskip}\noindent{\bfseries Proof.}\enspace\ignorespaces}{\par\addvspace{0.5\baselineskip}}
\renewcommand{\qed}{\relax\ifmmode\hskip2em \Box\else\unskip\nobreak\hskip1em $\Box$\fi}

\usepackage[most]{tcolorbox}
\tcbset{resstyle/.style={enhanced,breakable,colback=white,boxrule=0pt,frame hidden,arc=0pt,
  borderline west={1.6pt}{0pt}{black!55},left=7pt,right=6pt,top=3pt,bottom=3pt,before skip=9pt,after skip=9pt}}
\tcbset{defstyle/.style={enhanced,breakable,colback=white,boxrule=0pt,frame hidden,arc=0pt,
  borderline west={0.8pt}{0pt}{black!45},left=7pt,right=0pt,top=1pt,bottom=1pt,before skip=8pt,after skip=8pt}}
\tcolorboxenvironment{thm}{resstyle}
\tcolorboxenvironment{lemma}{resstyle}
\tcolorboxenvironment{proposition}{resstyle}
\tcolorboxenvironment{corollary}{resstyle}
\tcolorboxenvironment{definition}{defstyle}
\tcolorboxenvironment{assumption}{defstyle}
\titleformat{\section}{\large\bfseries}{\thesection}{0.8em}{}[\vspace{-0.6ex}\textcolor{black!35}{\rule{\textwidth}{0.4pt}}]

\begin{document}

\thispagestyle{first}
\begin{center}
{\small\itshape Preprint. Submitted to \emph{Automatica}. November 2026.\par}
\vspace{1.8em}
{\LARGE\bfseries Separating physics from discrepancy: Orthogonalized kernels for physics-encoded system identification\par}
\vspace{1.5em}
{\large Cesare Donati\textsuperscript{a,*}, Carlo Novara\textsuperscript{b}, Fabrizio Dabbene\textsuperscript{a}, Martina Mammarella\textsuperscript{a}\par}
\vspace{0.9em}
{\small\textsuperscript{a}\,Istituto di Elettronica e di Ingegneria dell'Informazione e delle Telecomunicazioni, Consiglio Nazionale delle Ricerche (CNR-IEIIT), c/o Politecnico di Torino, Corso Duca degli Abruzzi 24, 10129 Torino, Italy\par\smallskip
\textsuperscript{b}\,Dipartimento di Elettronica e Telecomunicazioni, Politecnico di Torino, Corso Duca degli Abruzzi 24, 10129 Torino, Italy\par}
\end{center}
\vspace{1.2em}
{\small\leftskip=1cm\rightskip=1cm\noindent\textbf{Abstract.}
Physics-encoded identification combines a physical model with a data-driven correction, but the data alone do not determine how the input--output map is split between them: components representable by the physical model may be assigned to either part. In kernel-based model embedding, this allocation is governed by the kernel.
As a consequence, the physical parameters lack a well-defined target. In this work, this ambiguity is resolved a priori by defining a canonical physical parameter as the parameter of the physical model that best approximates the noise-free map, in a weighted least-squares sense, over a user-defined operating domain, together with an orthogonal discrepancy.
Specifically, an orthogonalized kernel is derived that restricts the correction to functions orthogonal to the physical model. The resulting estimator keeps the closed-form solution of model embedding, while the canonical parameter depends only on the prescribed geometry, not on the kernel or the data.
As the inputs become dense and the noise vanishes, the set of parameters consistent with the data converges to the canonical parameter, a property that generally fails without orthogonalization. Under energy-bounded noise, suitable regularization attains optimal worst-case estimation accuracy. Numerical examples demonstrate consistent parameter estimates across different kernels and experiments.
\par
\medskip\noindent\textbf{Keywords:}
Kernel methods; nonlinear system identification; parameter estimation; physics-encoded identification; reproducing kernel Hilbert spaces; set-membership identification
\par}
\vspace{1.5em}
\begingroup\renewcommand{\thefootnote}{}\footnotetext{\hspace{-1.8em}\textsuperscript{*}\,Corresponding author: \href{mailto:cesaredonati@cnr.it}{cesaredonati@cnr.it}}\endgroup

\section{Introduction}
\label{sec:introduction}

In system identification, exploiting all available reliable physical {knowledge} can improve the generalization capability of {data-driven} models. Physical laws and equations encode relationships that remain valid beyond the observed data, enabling the identified model to generalize over operating conditions poorly represented in the training set. However, physical models are often incomplete due to simplifying assumptions, uncertain parameters, and neglected phenomena. In contrast, purely black-box models offer greater flexibility in capturing complex system behaviors but typically require large and sufficiently representative data sets and may generalize poorly beyond the observed operating conditions. Combining the structure provided by physical laws with the flexibility of a black-box parametrization offers a promising way to capture behaviors that the available physical description cannot adequately represent. \textit{Physics-encoded identification} follows this approach by combining a physics-based model with a data-driven correction term~\cite{donati2025kernel}, thereby explicitly embedding the physical knowledge in the model equations. This distinguishes it from \textit{physics-informed learning}, where physics is typically incorporated into the training loss of a black-box model~\cite{faroughi2024PhysicsNeuralNetworks}.

Within the physics-encoded framework, an additive formulation combines a physics-based component that captures known mechanisms with a flexible discrepancy term accounting for unmodeled effects. The two components can be estimated jointly, allowing the physical parameters to be calibrated while learning the remaining model mismatch. Kernel-based model embedding (KME)~\cite{donati2025kernel} provides a computationally convenient realization of this approach: by modeling the discrepancy in a reproducing kernel Hilbert space (RKHS), joint estimation admits a closed-form solution when the physical model is affine in its parameters.

However, the existence of a solution does not guarantee that the physical parameters and discrepancy can be uniquely distinguished. Indeed, the additive structure introduces an inherent ambiguity: any contribution of the discrepancy that can be reproduced by the physical model can be absorbed into the parameters without changing the input--output map. As a consequence, no data set can distinguish between these equivalent decompositions~\cite{KennedyOHagan,tuo2016theoretical}. This is particularly relevant when the physical parameters are of interest beyond predictive accuracy, as they may be interpreted, compared across experiments, or used for control and monitoring purposes. 

This ambiguity, inherent to the additive structure rather than to any particular estimator, has three main consequences. First, the physical parameters lack a uniquely defined target unless an additional rule is introduced to select a particular decomposition. Second, every estimator that returns a unique solution implicitly imposes such a rule. Least squares applied to the physical model alone yields a discrepancy \textit{orthogonal} to the physical regressors over the observed data, making the estimated parameters dependent on the input distribution and potentially different across experiments on the same system. In KME, the decomposition is instead governed by the regularization and kernel choice. Third, the estimated discrepancy cannot be unambiguously interpreted as a measure of model deficiency, since it may contain contributions that could equally be represented by the physical model.

The same non-uniqueness has been extensively studied in the calibration of computer models. In the Bayesian framework of Kennedy and O'Hagan~\cite{KennedyOHagan}, where the discrepancy is modeled as a Gaussian process, the calibration parameter is not uniquely determined by the input--output relationship. A natural way to resolve this ambiguity is to define the target parameter as the minimizer of the $L^2$ distance between the true response and the physical model, obtained by integrating their squared difference over the input domain. The work in~\cite{tuo2016theoretical} showed that, in general, a simplified version of the Kennedy--O'Hagan approach does not converge to the $L^2$ target as the data become dense. Instead, it converges to the parameter minimizing the RKHS norm of the residual, yielding a kernel-dependent target. This result was later extended to noisy observations with known covariance under suitable conditions~\cite{tuo2020improved}. 

To address this limitation, $L^2$ calibration~\cite{tuo2016theoretical,tuo2015efficient} estimates the target directly through a nonparametric fit followed by a projection onto the physical model, achieving semiparametric efficiency under Gaussian noise~\cite{tuo2015efficient}. The projected kernel approach of~\cite{tuo2019adjustments} provides an asymptotically equivalent alternative. In~\cite{plumlee2017bayesian}, the calibration parameter is defined by minimizing a possibly weighted loss, while the discrepancy is assigned a Gaussian process prior constrained to be orthogonal to the model gradient, extending the orthogonal Gaussian process formulation of~\cite{plumlee2018orthogonal}. The kernel derived in the present paper has the same structure as these covariance functions. Moreover, when the physical model is affine in the parameters, with a noninformative prior on them and fixed hyperparameters, the proposed estimator coincides with the posterior mean of~\cite{plumlee2017bayesian}. Despite this connection, the present work adopts a deterministic system-identification perspective, focusing on finite-data identifiability and worst-case estimation guarantees rather than statistical asymptotics and Bayesian uncertainty quantification.

Beyond the statistical calibration literature, the problem of separating physical and learned components has also been addressed in system identification through several regularization strategies. These include imposing sparsity on a correction term represented through basis functions~\cite{donati2025OffWhiteSparse}, penalizing deviations of the physical parameters from their least-squares estimate~\cite{bolderman2022feedforward}, penalizing the component of the network output that can be reproduced by the physical model on the training data~\cite{kon2022physics}, or introducing regularization within a linear-fractional representation that unifies different augmentation structures~\cite{hoekstra2025learning}. More recently, the approach in~\cite{gyorok2026orthogonal} eliminates the need for a trade-off weight by projecting the learning component onto the orthogonal complement of the physical regressors on the training data, independently of the chosen approximator. This projection separates the prediction-error criterion into two terms, only one of which depends on the physical parameters, and the physical estimate is the least-squares estimate of the physical model alone. Its target thus changes with the distribution of the data, and, without noise, it equals the true parameter only if the physical regressors are orthogonal to the discrepancy on the data. 

Motivated by these considerations, we develop an orthogonalized extension of the KME framework based on a user-defined reference geometry, specified by an operating domain and a suitable weight. The proposed framework preserves the computational tractability of joint estimation while providing a data- and kernel-independent definition of the physical parameter target. The main contributions are summarized as follows. First, we establish a \textit{canonical decomposition} of the input--output map into a physical component and an orthogonal discrepancy, uniquely identified by the reference geometry. Second, we derive an \textit{orthogonalized kernel} for any continuous kernel, restricting the associated RKHS to functions orthogonal to the physical model. Unlike sample-based orthogonalization, the resulting decomposition is defined at the function level. The estimator retains the closed-form structure of KME, requiring only a modification of the kernel matrix. Third, we establish deterministic identifiability guarantees within a set-membership framework. In particular, we show that the feasible parameter set converges to the canonical parameter as the inputs become dense and the noise vanishes, a property not generally guaranteed without orthogonalization. Fourth, under energy-bounded noise, we establish a worst-case accuracy bound and show that the proposed estimator achieves the optimal bound for a suitable choice of the regularization weight. We further show that an a posteriori $L^2$ projection of the KME estimate onto the physical model~\cite{tuo2019adjustments} need not enjoy the same guarantee. Finally, we provide closed-form expressions and quadrature rules for the required integrals. The results are illustrated on a static map and a simulated mass-spring-damper. In the latter, the mean estimates of the proposed estimator lie within $2\%$ of the canonical parameter across five experiments, and the median widths of the feasible sets along the stiffness and damping are over two orders of magnitude smaller than without orthogonalization; with Coulomb friction, which lies outside the RKHS, and the best regularization weight, the a posteriori $L^2$ projection of the KME fit~\cite{tuo2019adjustments} exceeds the optimal bound by up to a factor of $3.8$ (median over data sets), whereas the proposed estimator attains it. Orthogonalization thus resolves the ambiguity of the additive decomposition: the physical parameters acquire a unique target, fixed before estimation, while the computational structure of KME and a comparable prediction accuracy are retained.

\subsubsection*{Outline} The remainder of the paper is organized as follows. Section~\ref{sec:kerneltheory} recalls kernel-based model embedding. Section~\ref{sec:orthker} introduces the reference geometry, the canonical parameter, the orthogonalized kernel, and the resulting estimator. Section~\ref{sec:sm} presents the set-membership analysis. Section~\ref{sec:impl} addresses the implementation, and Section~\ref{sec:num} reports the numerical results. Section~\ref{sec:concl} concludes the paper, and the Supplementary material details the computation of the moment representers.

\subsubsection*{Notation} $\mathbf{I}_T$ is the $T\times T$ identity matrix; for symmetric matrices, $A\succeq0$ ($A\succ0$) denotes positive semidefiniteness (definiteness), $A\preceq B$ means $B-A\succeq0$, $\lambda_{\min}(A)$ is the smallest eigenvalue of $A$, and, for $A\succeq0$, $A^{1/2}$ is its positive semidefinite square root; $A^\dagger$ is the Moore--Penrose pseudoinverse of $A$; $\|\cdot\|_1$, $\|\cdot\|_2$, and $\|\cdot\|_\infty$ are the $\ell_1$, the Euclidean, and the maximum norm of a vector, respectively.

\section{Kernel-based model embedding}
\label{sec:kerneltheory}
This section reviews the kernel-based model embedding framework introduced in~\cite{donati2025kernel}.

\begin{definition}[positive definite kernel]\label{def:pdkernel}
Let $\mathcal{X}$ be a nonempty set. A real-valued, continuous, symmetric function $\kappa:\mathcal{X}\times\mathcal{X}\to\mathbb{R}$ is a positive definite kernel on $\mathcal{X}$ if $\sum_{i=1}^n\sum_{j=1}^n c_ic_j\kappa(x_i,x_j)\geq0$ for all $n\in\mathbb{N}$, $x_1,\dots,x_n\in\mathcal{X}$, and $c_1,\dots,c_n\in\mathbb{R}$.
\end{definition}

\begin{definition}[reproducing kernel Hilbert space]\label{def:RKHS}
Let $\mathcal{H}$ be a Hilbert space of real-valued functions on a nonempty set $\mathcal{X}$, with inner product $\langle\cdot,\cdot\rangle_\mathcal{H}$ and norm $\|\cdot\|_\mathcal{H}$. A function $\kappa:\mathcal{X}\times\mathcal{X}\to\mathbb{R}$ is a reproducing kernel of $\mathcal{H}$, and $\mathcal{H}$ is a reproducing kernel Hilbert space (RKHS) on $\mathcal{X}$, if (i) $\kappa(\cdot,x)\in\mathcal{H}$ for all $x\in\mathcal{X}$, and (ii) $\langle h,\kappa(\cdot,x)\rangle_\mathcal{H}=h(x)$ for all $x\in\mathcal{X}$ and $h\in\mathcal{H}$ (reproducing property).
\end{definition}

By the Moore--Aronszajn theorem~\cite{aronszajn1950theory}, every positive definite kernel $\kappa$ is the reproducing kernel of a unique RKHS $\mathcal{H}$, and $\kappa(x,x')=\langle\kappa(\cdot,x),\kappa(\cdot,x')\rangle_\mathcal{H}$. Given $x_1,\dots,x_T\in\mathcal{X}$, the kernel matrix $\mathbf{K}\in\mathbb{R}^{T\times T}$, with entries $\mathbf{K}_{ij}\doteq\kappa(x_i,x_j)$, satisfies $\mathbf{K}\succeq0$ by Definition~\ref{def:pdkernel}.

Consider data $\mathcal{D}=\{(x_1,y_1),\dots,(x_T,y_T)\}$ generated by the map
\begin{equation}
y=f(x)^\top\bar\theta+\Delta(x)+e,
\label{eqn:system.comp.lin}
\end{equation}
where $f:\mathcal{X}\to\mathbb{R}^{n_\theta}$ is a known basis, derived, e.g., from physical principles, $\bar\theta\in\mathbb{R}^{n_\theta}$ collects the physical parameters, $\Delta:\mathcal{X}\to\mathbb{R}$ is an unknown discrepancy accounting for, e.g., modeling errors, and $e$ is an error term; a known additive term without parameters can be subtracted from the outputs and is therefore omitted. The kernel-based model embedding (KME) problem of~\cite{donati2025kernel} estimates the physical parameters and a nonparametric correction $\delta\in\mathcal{H}$ jointly,
\begin{equation}
(\theta^\star,\delta^\star)=\arg\!\!\min_{\theta\in\mathbb{R}^{n_\theta},\,\delta\in\mathcal{H}}\sum_{t=1}^{T}\bigl[y_t-\hat y_t(\theta,\delta)\bigr]^2+\gamma\|\delta\|^2_\mathcal{H},
\label{eqn:optprob.theta}
\end{equation}
with $\hat y_t(\theta,\delta)\doteq f(x_t)^\top\theta+\delta(x_t)$ and $\gamma>0$. In~\cite{donati2025kernel}, where nonlinear parametrizations are also treated, $\Delta$ is assumed to belong to $\mathcal{H}$, so that the correction can represent it exactly. Let
\begin{equation}
F(x)\doteq\begin{bmatrix}f(x_1)^\top\\ \vdots\\ f(x_T)^\top\end{bmatrix},
\quad
Y\doteq\begin{bmatrix}y_1\\ \vdots\\ y_T\end{bmatrix},
\quad
\Psi\doteq(\mathbf{K}+\gamma\mathbf{I}_T)^{-1}.
\label{eqn:Psidef}
\end{equation}
For fixed $\theta$, the minimization over $\delta$ in~\eqref{eqn:optprob.theta} is a kernel ridge regression problem with outputs $Y-F(x)\theta$, whose solution is a combination of the kernel sections $\kappa(\cdot,x_j)$ by the representer theorem~\cite{scholkopf2001.reprthm,saunders1998KRR}; substituting it yields a weighted least-squares problem in $\theta$. If $F(x)$ has full column rank, the unique solution of~\eqref{eqn:optprob.theta} is~\cite[Theorems~1 and~2]{donati2025kernel}
\begin{subequations}
\begin{align}
\theta^\star&=\bigl(F(x)^\top\Psi F(x)\bigr)^{-1}F(x)^\top\Psi Y,
\label{eqn:thmlincf}\\
\delta^\star(x)&=\textstyle\sum_{j=1}^{T}\omega_j^\star\kappa(x,x_j),
\quad
\omega^\star\doteq\Psi\bigl(Y-F(x)\theta^\star\bigr).
\label{eqn:reprthm.ext}
\end{align}
\end{subequations}
Since $\mathbf{K}\succeq0$ and $\gamma>0$, $\Psi\succ0$; hence, $F(x)^\top\Psi F(x)$ is invertible if and only if $F(x)$ has full column rank, which requires $T\geq n_\theta$ and sufficiently exciting inputs.

\section{Orthogonal kernel-based model embedding}\label{sec:orthker}
 
Throughout this section, $f=[f_1,\dots,f_{n_\theta}]^\top$ and $\bar g\doteq f^\top\bar\theta+\Delta$ denotes the noise-free map. First we note that the pair $(\bar\theta,\Delta)$ is not determined by $\bar g$: for every $a\in\mathbb{R}^{n_\theta}$,
\begin{equation}
\bar g=f^\top(\bar\theta+a)+\bigl(\Delta-f^\top a\bigr),
\label{eqn:ok.ambiguity}
\end{equation}
and the two decompositions produce the same output at every $x\in\mathcal{X}$, so that no data set can distinguish them~\cite{KennedyOHagan,tuo2016theoretical}. Hence, the physical parameter has no well-defined target unless a rule selecting one member of \eqref{eqn:ok.ambiguity} is prescribed. In Problem~\eqref{eqn:optprob.theta}, the selection is made implicitly by the regularizer, and therefore depends on the kernel and on its hyperparameters: different kernels may yield different physical parameters with the same fitted map. Moreover, part of the estimated discrepancy may describe behavior that the physical model can already reproduce. In this section, a selection rule is defined through a reference geometry, independently of the estimator, and a kernel is constructed whose correction space complies with it.
 
\subsection{Physical span and reference geometry}\label{sec:ok.geometry}
 
Let $\mathcal{X}\subseteq\mathbb{R}^{n_x}$, and let $\mathcal{F}\doteq\operatorname{span}\{f_1,\dots,f_{n_\theta}\}$ denote the \emph{physical span}, i.e., the linear space of functions $f^\top c:\mathcal{X}\to\mathbb{R}$, $c\in\mathbb{R}^{n_\theta}$, that the physical model can represent by a suitable choice of its parameters.

Selecting one member of \eqref{eqn:ok.ambiguity} requires stating when a discrepancy contains no component that the physical model could explain, i.e., a notion of orthogonality between functions. We define it through two modeling choices: an \emph{operating domain} $\Omega\subseteq\mathcal{X}$, i.e., the region of the input space where the physical model is to be interpreted, and a \emph{reference weight} $\rho:\Omega\to[0,\infty)$, which assigns a relative importance to the points of $\Omega$. Both are chosen by the user, and, together, they define the \emph{reference inner product} and norm
\begin{equation}
\langle g,q\rangle_\rho\doteq\int_\Omega g(x)q(x)\rho(x)\,dx,
\quad
\|g\|_\rho\doteq\langle g,g\rangle_\rho^{1/2},
\label{eqn:ok.inner}
\end{equation}
and the corresponding Gram matrix of the physical basis,
\begin{equation}
G\doteq\int_\Omega f(x)f(x)^\top\rho(x)\,dx\in\mathbb{R}^{n_\theta\times n_\theta},
\label{eqn:ok.G}
\end{equation}
whose entries are $G_{ij}=\langle f_i,f_j\rangle_\rho$. The following assumption ensures that these quantities are well defined and that the physical basis is nondegenerate in this geometry.

\begin{assumption}[reference geometry]\label{ass:ok.geometry}
(i) $\Omega$ is compact; (ii) $\rho$ is integrable on $\Omega$; (iii) each $f_j$ is continuous on $\Omega$; (iv) $G\succ0$.
\end{assumption}

Items \textit{(i)}--\textit{(iii)} are mild: they hold for any bounded and closed operating region, any bounded weight, including the uniform choice $\rho\equiv1$, and any continuous basis. Continuity of the $f_j$ can be relaxed to boundedness and piecewise continuity, i.e., continuity outside a set of zero measure. Item \textit{(iv)} requires the basis functions to be linearly independent, {in the sense that no nonzero combination $f^\top c$ vanishes almost everywhere
on $\{x\in\Omega:\rho(x)>0\}$.}
If it fails, some combination of the physical parameters has no effect on the model over the region of interest, and should be removed by reparametrization. Item \textit{(iv)} is thus the counterpart, on $\Omega$, of the full-column-rank condition on $F(x)$ recalled after \eqref{eqn:thmlincf}.

We denote by $L^2_\rho$ the {space} of functions $g$ on $\Omega$ with $\|g\|_\rho<\infty$, and write $g\perp_\rho\mathcal{F}$ if $\langle f_j,g\rangle_\rho=0$ for $j=1,\dots,n_\theta$.

\begin{definition}[reference projection]\label{def:ok.proj}
For $g\in L^2_\rho$, the \emph{physical coefficient vector} of $g$ and its \emph{projection} onto $\mathcal{F}$ are

\begin{subequations}
\begin{align}\label{eqn:ok.proj}
\vartheta_\rho(g)&\doteq\arg\min_{c\in\mathbb{R}^{n_\theta}}\|g-f^\top c\|_\rho^2,\\
\Pi_\rho g&\doteq f^\top\vartheta_\rho(g).
\end{align}
\end{subequations}
\end{definition}

The cost in \eqref{eqn:ok.proj} is a quadratic function of $c$ with Hessian $2G\succ0$; hence, the minimizer exists and is unique. Thus, $\Pi_\rho g$ is the best approximation of $g$ by the physical model over $\Omega$, in the sense of the reference norm, and $\vartheta_\rho(g)$ collects the corresponding parameters. Note that $\Pi_\rho$ depends on the physical model only through its span $\mathcal{F}$: replacing $f$ with $Af$, $A$ nonsingular, leaves $\Pi_\rho g$ unchanged and maps $\vartheta_\rho(g)$ into $A^{-\top}\vartheta_\rho(g)$. The following properties of the reference projection are used throughout.

\begin{proposition}[reference projection]\label{prop:ok.proj}
Let Assumption~\ref{ass:ok.geometry} hold and let $g\in L^2_\rho$. Then:
\begin{itemize}
\item[(i)] the minimizer in \eqref{eqn:ok.proj} is
\begin{equation}
\vartheta_\rho(g)=G^{-1}\!\int_\Omega f(x)g(x)\rho(x)\,dx ;
\label{eqn:ok.projcf}
\end{equation}
\item[(ii)] $\vartheta_\rho$ is linear in $g$, and $\vartheta_\rho(f^\top c)=c$ for all $c\in\mathbb{R}^{n_\theta}$;
\item[(iii)] $(g-\Pi_\rho g)\perp_\rho\mathcal{F}$;
\item[(iv)] $\|\vartheta_\rho(g)\|_2\leq\lambda_{\min}(G)^{-1/2}\,\|g\|_\rho$.
\end{itemize}
\end{proposition}

\begin{pf}
Let $b(g)\doteq\int_\Omega f(x)g(x)\rho(x)\,dx$, whose $j$-th entry is $\langle f_j,g\rangle_\rho$ and is finite by the Cauchy--Schwarz inequality, since $f_j$ is bounded on $\Omega$ and $\rho$ is integrable.

\textit{(i)} The cost in \eqref{eqn:ok.proj} equals $\|g\|_\rho^2-2c^\top b(g)+c^\top Gc$, a quadratic function of $c$ with gradient $2(Gc-b(g))$; since $G\succ0$, its unique minimizer is $c=G^{-1}b(g)$, i.e., \eqref{eqn:ok.projcf}.

\textit{(ii)} The map $b$ is linear, and so is $\vartheta_\rho=G^{-1}b$. Moreover, $b(f^\top c)=Gc$ by \eqref{eqn:ok.G}, so that $\vartheta_\rho(f^\top c)=c$.

\textit{(iii)} Since $\Pi_\rho g=f^\top\vartheta_\rho(g)$, (ii) and (i) give $b(\Pi_\rho g)=G\vartheta_\rho(g)=b(g)$. Hence, by linearity, $b(g-\Pi_\rho g)=0$, i.e., $\langle f_j,g-\Pi_\rho g\rangle_\rho=0$ for $j=1,\dots,n_\theta$, which is (iii).

\textit{(iv)} By (iii) and linearity of the inner product, $g-\Pi_\rho g$ is $\rho$-orthogonal to $\Pi_\rho g\in\mathcal{F}$. Hence, by the Pythagorean identity and \eqref{eqn:ok.G},
\[
\begin{aligned}
\|g\|_\rho^2&=\|\Pi_\rho g\|_\rho^2+\|g-\Pi_\rho g\|_\rho^2\geq\|\Pi_\rho g\|_\rho^2\\
&=\vartheta_\rho(g)^\top G\,\vartheta_\rho(g)\geq\lambda_{\min}(G)\,\|\vartheta_\rho(g)\|_2^2,
\end{aligned}
\]
and (iv) follows by dividing by $\lambda_{\min}(G)>0$ and taking square roots. \qed
\end{pf}

By Proposition~\ref{prop:ok.proj}, $\Pi_\rho$ is the $\langle\cdot,\cdot\rangle_\rho$-orthogonal projector onto $\mathcal{F}$, and $\vartheta_\rho(g)$ collects the coordinates of $\Pi_\rho g$ in the basis $f_1,\dots,f_{n_\theta}$.
 
\subsection{Canonical physical parameter and orthogonal discrepancy}\label{sec:ok.canonical}
 
The reference geometry provides a rule for selecting one member of \eqref{eqn:ok.ambiguity}. The physical parameter is taken as the coefficient vector of the reference projection of $\bar g$ onto $\mathcal{F}$, and the discrepancy as the corresponding residual. As shown next, this is the only member of \eqref{eqn:ok.ambiguity} whose discrepancy is $\rho$-orthogonal to $\mathcal{F}$, i.e., contains no component that the physical model can reproduce on $\Omega$.
 
\begin{definition}[canonical parameter]\label{def:ok.canonical}
Let Assumption~\ref{ass:ok.geometry} hold and let the restriction of $\Delta$ to $\Omega$ belong to $L^2_\rho$. The {canonical physical parameter} $\theta_\rho$, the {discrepancy coefficient} $\theta_\Delta$, and the {orthogonal discrepancy} $\Delta_\perp$ are
\begin{equation}
\theta_\rho\doteq\vartheta_\rho(\bar g),
\quad
\theta_\Delta\doteq\vartheta_\rho(\Delta),
\quad
\Delta_\perp(x)\doteq\Delta(x)-f(x)^\top\theta_\Delta,
\label{eqn:ok.canonical}
\end{equation}
for all $x\in\mathcal{X}$.
\end{definition}
 
\begin{proposition}[orthogonal decomposition]\label{prop:ok.decomp}
Under the assumptions of Definition~\ref{def:ok.canonical}: (i) $\theta_\rho=\bar\theta+\theta_\Delta$; (ii) the map \eqref{eqn:system.comp.lin} can be written, for all $x\in\mathcal{X}$, as
\begin{equation}
y=f(x)^\top\theta_\rho+\Delta_\perp(x)+e,
\qquad
\Delta_\perp\perp_\rho\mathcal{F};
\label{eqn:ok.decomp}
\end{equation}
(iii) $(\theta_\rho,\Delta_\perp)$ is the only member of the family \eqref{eqn:ok.ambiguity} whose discrepancy is $\rho$-orthogonal to $\mathcal{F}$.
\end{proposition}
\begin{pf}
\textit{(i)} follows from linearity of $\vartheta_\rho$ and $\vartheta_\rho(f^\top\bar\theta)=\bar\theta$:
\[
\theta_\rho=\vartheta_\rho(f^\top\bar\theta+\Delta)=\vartheta_\rho(f^\top\bar\theta)+\vartheta_\rho(\Delta)=\bar\theta+\theta_\Delta .
\]
\textit{(ii)} follows by adding and subtracting $f^\top\theta_\Delta$ in \eqref{eqn:system.comp.lin},
\[
y=f(x)^\top(\bar\theta+\theta_\Delta)+\bigl(\Delta(x)-f(x)^\top\theta_\Delta\bigr)+e,
\]
and using (i). Moreover, $\Delta_\perp=\Delta-f^\top\vartheta_\rho(\Delta)=\Delta-\Pi_\rho\Delta$, so that $\Delta_\perp\perp_\rho\mathcal{F}$ by Proposition~\ref{prop:ok.proj}(iii).
\textit{(iii)} The members of \eqref{eqn:ok.ambiguity} are the pairs $(\bar\theta+a,\Delta-f^\top a)$, $a\in\mathbb{R}^{n_\theta}$. For $g\in L^2_\rho$, \eqref{eqn:ok.projcf} gives $G\vartheta_\rho(g)=\int_\Omega f(x)g(x)\rho(x)\,dx$, whose $j$-th entry is $\langle f_j,g\rangle_\rho$; since $G\succ0$, it follows that $g\perp_\rho\mathcal{F}$ if and only if $\vartheta_\rho(g)=0$. Moreover, by Proposition~\ref{prop:ok.proj}(ii),
\[
\vartheta_\rho(\Delta-f^\top a)=\vartheta_\rho(\Delta)-\vartheta_\rho(f^\top a)=\theta_\Delta-a .
\]
Hence, $\Delta-f^\top a\perp_\rho\mathcal{F}$ if and only if $a=\theta_\Delta$, and the corresponding member is $(\bar\theta+\theta_\Delta,\Delta-f^\top\theta_\Delta)=(\theta_\rho,\Delta_\perp)$, by (i) and \eqref{eqn:ok.canonical}. \qed
\end{pf}
 
By Proposition~\ref{prop:ok.decomp}, $\Delta=f^\top\theta_\Delta+\Delta_\perp$, where $f^\top\theta_\Delta=\Pi_\rho\Delta\in\mathcal{F}$ is reproduced exactly by moving the parameters from $\bar\theta$ to $\theta_\rho$, whereas $\Delta_\perp$, for which $\vartheta_\rho(\Delta_\perp)=0$, describes the behavior that the physical model cannot reproduce on $\Omega$. The canonical parameter, i.e., the coefficient vector of the best approximation of $\bar g$ in $\mathcal{F}$, is taken as the target of identification in the following. It depends only on $(\bar g,f,\Omega,\rho)$: no data set, kernel, regularization weight, or estimator enters its definition, and estimates obtained with different kernels or from different experiments refer to the same parameter whenever the same {physical parametrization $f$ and} reference geometry are adopted. Moreover, $\theta_\rho=\bar\theta$ if and only if $\Delta\perp_\rho\mathcal{F}$. Otherwise, the difference $\theta_\Delta$ is not an estimation error, since the members of \eqref{eqn:ok.ambiguity} cannot be distinguished from $\bar g$, and it is meaningful only if $\bar\theta$ has an independent definition, e.g., from first principles or separate experiments. In this sense, Proposition~\ref{prop:ok.decomp} is a normalization rather than an identifiability result, as it fixes which parameter is regarded as physical.

\begin{remark}[decomposition selected by KME]\label{rem:ok.implicit}
{In} the absence of a prescribed rule, the selection in \eqref{eqn:ok.ambiguity}
{is made by the KME problem} \eqref{eqn:optprob.theta}. Let $(\theta^\star,\delta^\star)$ solve \eqref{eqn:optprob.theta}, {let
$\mathcal{V}\doteq\mathcal{F}\cap\mathcal{H}$ denote the physical functions that are
also admissible corrections, and let $u=f^\top a\in\mathcal{V}$. It is readily verified that,}
for every $s\in\mathbb{R}$, the pair $(\theta^\star+sa,\delta^\star-su)$ yields the same predictions as $(\theta^\star,\delta^\star)$.
{Hence, optimality of $(\theta^\star,\delta^\star)$}
implies that
\[
\|\delta^\star\|_\mathcal{H}^2\leq\|\delta^\star-su\|_\mathcal{H}^2,
\]
i.e., $0\leq-2s\langle\delta^\star,u\rangle_\mathcal{H}+s^2\|u\|_\mathcal{H}^2$ for all $s\in\mathbb{R}$. Since the right-hand side vanishes at $s=0$, this requires its derivative at $s=0$ to vanish, i.e.,
\[
\langle\delta^\star,u\rangle_\mathcal{H}=0.
\]
Thus, $\delta^\star$ is \emph{orthogonal to} {$\mathcal{V}$}, {\emph{but in the geometry of}} $\mathcal{H}${: this orthogonality is dictated by $\kappa$ and its hyperparameters rather than prescribed by the user, and in general KME does not target $\theta_\rho$.}
\end{remark}

{The previous remark also suggests the remedy: if the admissible corrections are
restricted to functions that are $\rho$-orthogonal to $\mathcal{F}$, the canonical
decomposition is the only admissible one, and nothing is left to the geometry of the
kernel. The next subsection constructs a kernel whose RKHS is exactly this set.}

\subsection{Orthogonalized kernel}\label{sec:ok.kernel}
 
The prescription of Section~\ref{sec:ok.canonical} is enforced on the correction by restricting it to the set
\begin{equation}
\mathcal{H}_\perp\doteq\bigl\{\delta\in\mathcal{H}:\ \langle f_j,\delta\rangle_\rho=0,\ j=1,\dots,n_\theta\bigr\}.
\label{eqn:ok.Hperp}
\end{equation}
The aim is a kernel whose RKHS is $\mathcal{H}_\perp$. Since the reproducing kernel of a closed subspace of $\mathcal{H}$ is obtained by projecting the kernel sections $\kappa(\cdot,x)$ onto it with the $\langle\cdot,\cdot\rangle_\mathcal{H}$-orthogonal projector, the constraints in \eqref{eqn:ok.Hperp}, stated in the reference geometry, are first rewritten as orthogonality conditions in $\mathcal{H}$, under the following assumption.
 
\begin{assumption}[kernel]\label{ass:ok.kernel}
The kernel $\kappa$ satisfies Definition~\ref{def:pdkernel} on $\mathcal{X}$.
\end{assumption}
 
Under Assumptions~\ref{ass:ok.geometry} and~\ref{ass:ok.kernel}, every $\delta\in\mathcal{H}$ has two properties that are used below. First, by the reproducing property, $\delta(x)=\langle\delta,\kappa(\cdot,x)\rangle_\mathcal{H}$ and $\|\kappa(\cdot,x)\|_\mathcal{H}^2=\langle\kappa(\cdot,x),\kappa(\cdot,x)\rangle_\mathcal{H}=\kappa(x,x)$; hence, by the Cauchy--Schwarz inequality,
\begin{equation}
|\delta(x)|\leq\|\delta\|_\mathcal{H}\,\kappa(x,x)^{1/2},
\qquad x\in\mathcal{X},
\label{eqn:ok.pointbound}
\end{equation}
i.e., point values of $\delta$ are controlled by its RKHS norm. Second, $\delta$ is continuous, since, in the same way, $|\delta(x)-\delta(x')|\leq\|\delta\|_\mathcal{H}\|\kappa(\cdot,x)-\kappa(\cdot,x')\|_\mathcal{H}$, where $\|\kappa(\cdot,x)-\kappa(\cdot,x')\|_\mathcal{H}^2=\kappa(x,x)-2\kappa(x,x')+\kappa(x',x')\to0$ as $x'\to x$. Squaring \eqref{eqn:ok.pointbound}, multiplying by $\rho$, and integrating over $\Omega$ yields
\begin{equation}
\|\delta\|_\rho^2\leq c_\kappa\|\delta\|_\mathcal{H}^2,
\qquad
c_\kappa\doteq\int_\Omega\kappa(x,x)\rho(x)\,dx,
\label{eqn:ok.embed}
\end{equation}
where $c_\kappa<\infty$ because $\kappa$ is continuous, and hence bounded, on the compact set $\Omega$, and $\rho$ is integrable.
 
\begin{lemma}[moment representers]\label{lem:ok.riesz}
Let Assumptions~\ref{ass:ok.geometry} and~\ref{ass:ok.kernel} hold, and define
\begin{equation}
\eta(x)\doteq\int_\Omega f(\xi)\,\kappa(\xi,x)\,\rho(\xi)\,d\xi\in\mathbb{R}^{n_\theta},
\qquad x\in\mathcal{X}.
\label{eqn:ok.eta}
\end{equation}
Then, each component $\eta_j$ belongs to $\mathcal{H}$ and
\begin{equation}
\langle\delta,\eta_j\rangle_\mathcal{H}=\langle f_j,\delta\rangle_\rho,
\qquad \forall\,\delta\in\mathcal{H},\ j=1,\dots,n_\theta.
\label{eqn:ok.riesz}
\end{equation}
Moreover, the Gram matrix $H\doteq[\langle\eta_i,\eta_j\rangle_\mathcal{H}]_{i,j=1}^{n_\theta}$ is symmetric positive semidefinite and satisfies
\begin{equation}
H=\int_\Omega\!\int_\Omega f(x)\,\kappa(x,\xi)\,f(\xi)^\top\rho(x)\rho(\xi)\,d\xi\,dx.
\label{eqn:ok.H}
\end{equation}
\end{lemma}
\begin{pf}
For $\delta\in\mathcal{H}$, the Cauchy--Schwarz inequality for $\langle\cdot,\cdot\rangle_\rho$ and \eqref{eqn:ok.embed} yield
\[
|\langle f_j,\delta\rangle_\rho|\leq\|f_j\|_\rho\, c_\kappa^{1/2}\|\delta\|_\mathcal{H}.
\]
Thus, $\delta\mapsto\langle f_j,\delta\rangle_\rho$ is a bounded linear functional on $\mathcal{H}$ and, by the Riesz representation theorem~\cite[Sec.~5.3, Theorem~2]{luenberger1969optimization}, there exists a unique $\eta_j\in\mathcal{H}$ satisfying \eqref{eqn:ok.riesz}. Choosing $\delta=\kappa(\cdot,x)$ in \eqref{eqn:ok.riesz} and using the reproducing property of Definition~\ref{def:RKHS} gives
\[
\eta_j(x)=\langle\kappa(\cdot,x),\eta_j\rangle_\mathcal{H}=\langle f_j,\kappa(\cdot,x)\rangle_\rho,
\]
i.e., \eqref{eqn:ok.eta}. Finally, \eqref{eqn:ok.riesz} with $\delta=\eta_j$ gives
$\langle\eta_i,\eta_j\rangle_\mathcal{H}=\langle f_i,\eta_j\rangle_\rho,$
which is \eqref{eqn:ok.H}; the double integral is absolutely convergent, since, by the Cauchy--Schwarz inequality in~$\mathcal{H}$,
\[
|f_i(x)\kappa(x,\xi)f_j(\xi)|\leq|f_i(x)|\kappa(x,x)^{1/2}\kappa(\xi,\xi)^{1/2}|f_j(\xi)|,
\]
and the right-hand side is integrable on $\Omega\times\Omega$ with respect to $\rho(x)\rho(\xi)\,dx\,d\xi$. \qed
\end{pf}

By \eqref{eqn:ok.riesz}, $\delta\in\mathcal{H}$ satisfies the constraints in \eqref{eqn:ok.Hperp} if and only if $\langle\delta,\eta_j\rangle_\mathcal{H}=0$ for $j=1,\dots,n_\theta$, i.e., $\mathcal{H}_\perp$ is the orthogonal complement in $\mathcal{H}$ of $\operatorname{span}\{\eta_1,\dots,\eta_{n_\theta}\}$. Its reproducing kernel is thus obtained by removing from each $\kappa(\cdot,x')$ its $\mathcal{H}$-orthogonal projection onto this span. As in Definition~\ref{def:ok.proj}, with $\langle\cdot,\cdot\rangle_\rho$, $f$, and $G$ replaced by $\langle\cdot,\cdot\rangle_\mathcal{H}$, $\eta$, and $H$, this projection is $\eta^\top c$, where $c$ solves $Hc=\eta(x')$, the right-hand side following from the reproducing property. The minimum-norm solution $c=H^\dagger\eta(x')$, which also covers linearly dependent $\eta_j$, leads to the following definition.
 
\begin{definition}[orthogonalized kernel]\label{def:ok.kperp}
Under Assumptions~\ref{ass:ok.geometry} and~\ref{ass:ok.kernel}, the \emph{orthogonalized kernel} is
\begin{equation}
\kappa_\perp(x,x')\doteq\kappa(x,x')-\eta(x)^\top H^\dagger\eta(x'),
\qquad x,x'\in\mathcal{X},
\label{eqn:ok.kperp}
\end{equation}
where $H^\dagger$ is the Moore--Penrose pseudoinverse of $H$, and $H^\dagger=H^{-1}$ whenever $H\succ0$.
\end{definition}
 
\begin{thm}[orthogonalized kernel]\label{thm:ok.kperp}
Let Assumptions~\ref{ass:ok.geometry} and~\ref{ass:ok.kernel} hold. Then:
\begin{itemize}
\item[(i)] $\kappa_\perp$ satisfies Definition~\ref{def:pdkernel}, and $\kappa-\kappa_\perp$ is also a positive definite kernel; in particular, $0\leq\kappa_\perp(x,x)\leq\kappa(x,x)$;
\item[(ii)] $\mathcal{H}_\perp$ is a closed subspace of $\mathcal{H}$, and $\kappa_\perp(\cdot,x)=P_\perp\kappa(\cdot,x)$ for all $x\in\mathcal{X}$, where $P_\perp$ denotes the $\langle\cdot,\cdot\rangle_\mathcal{H}$-orthogonal projector onto $\mathcal{H}_\perp$;
\item[(iii)] $\mathcal{H}_\perp$, equipped with the restriction of $\langle\cdot,\cdot\rangle_\mathcal{H}$, is the RKHS associated with $\kappa_\perp$;
\item[(iv)] $\delta\in\mathcal{H}$ belongs to $\mathcal{H}_\perp$ if and only if $\Pi_\rho\delta=0$. In particular, $\kappa_\perp(\cdot,x)\perp_\rho\mathcal{F}$ for all $x\in\mathcal{X}$, and so is every finite combination $\sum_{j}\omega_j\kappa_\perp(\cdot,x_j)$;
\item[(v)] if $f^\top\theta_1+\delta_1=f^\top\theta_2+\delta_2$ with $\theta_1,\theta_2\in\mathbb{R}^{n_\theta}$ and $\delta_1,\delta_2\in\mathcal{H}_\perp$, then $\theta_1=\theta_2$ and $\delta_1=\delta_2$.
\end{itemize}
\end{thm}
 
\begin{pf}
Let
\[
b(\delta)\doteq\int_\Omega f(x)\delta(x)\rho(x)\,dx\in\mathbb{R}^{n_\theta}
\]
denote the vector of physical moments of $\delta\in\mathcal{H}$, as in the proof of Proposition~\ref{prop:ok.proj}. By \eqref{eqn:ok.riesz}, its entries are
\begin{equation}
b_j(\delta)=\langle\eta_j,\delta\rangle_\mathcal{H},
\label{eqn:ok.pf.bH}
\end{equation}
so that $\mathcal{H}_\perp=\{\delta\in\mathcal{H}:b(\delta)=0\}$. Two special cases are used repeatedly: by the reproducing property,
\begin{equation}
b(\kappa(\cdot,x))=\eta(x);
\label{eqn:ok.pf.bk}
\end{equation}
by \eqref{eqn:ok.riesz} and the definition of $H$, $b_i(\eta^\top c)=\langle\eta_i,\eta^\top c\rangle_\mathcal{H}=\sum_{j}H_{ij}c_j$, i.e.,
\begin{equation}
b(\eta^\top c)=Hc\quad\text{for every } c\in\mathbb{R}^{n_\theta}.
\label{eqn:ok.pf.beta}
\end{equation}

\emph{Step~1: $b(\delta)$ lies in the range of $H$.} If $c\in\ker H$, then $\|\eta^\top c\|_\mathcal{H}^2=c^\top Hc=0$, i.e., $\eta^\top c=0$, and hence, by \eqref{eqn:ok.pf.bH}, $c^\top b(\delta)=\langle\eta^\top c,\delta\rangle_\mathcal{H}=0$. Thus, $b(\delta)$ is orthogonal to $\ker H$ and, $H$ being symmetric, belongs to its range. Consequently,
\begin{equation}
HH^\dagger b(\delta)=b(\delta).
\label{eqn:ok.pf.range}
\end{equation}
If $H\succ0$, the last step is trivial.

\emph{Step~2: the projection onto $\mathcal{H}_\perp$.} For $\delta\in\mathcal{H}$, define
\begin{equation}
P_\perp\delta\doteq\delta-\eta^\top H^\dagger b(\delta).
\label{eqn:ok.pf.Pperp}
\end{equation}
By linearity of $b$, \eqref{eqn:ok.pf.beta} with $c=H^\dagger b(\delta)$, and \eqref{eqn:ok.pf.range},
\[
b(P_\perp\delta)=b(\delta)-HH^\dagger b(\delta)=0,
\]
i.e., $P_\perp\delta\in\mathcal{H}_\perp$. Moreover, $\delta-P_\perp\delta$ is a combination of $\eta_1,\dots,\eta_{n_\theta}$, and every $\eta_j$ is $\mathcal{H}$-orthogonal to $\mathcal{H}_\perp$, since, by \eqref{eqn:ok.pf.bH}, $\langle g,\eta_j\rangle_\mathcal{H}=b_j(g)=0$ for $g\in\mathcal{H}_\perp$. Therefore, $\delta=P_\perp\delta+(\delta-P_\perp\delta)$ splits $\delta$ into a component in $\mathcal{H}_\perp$ and a component $\mathcal{H}$-orthogonal to it, i.e., $P_\perp$ is the $\mathcal{H}$-orthogonal projector onto $\mathcal{H}_\perp$. Finally, $\mathcal{H}_\perp$ is closed, being the set where the continuous linear maps $b_1,\dots,b_{n_\theta}$ vanish.

\emph{Proof of (ii).} Applying \eqref{eqn:ok.pf.Pperp} to $\delta=\kappa(\cdot,x)$ and using \eqref{eqn:ok.pf.bk} gives
\[
P_\perp\kappa(\cdot,x)=\kappa(\cdot,x)-\eta(\cdot)^\top H^\dagger\eta(x)=\kappa_\perp(\cdot,x).
\]

\emph{Proof of (iii).} By (ii) and Step~2, $\kappa_\perp(\cdot,x)\in\mathcal{H}_\perp$. For every $\delta\in\mathcal{H}_\perp$, since $\delta$ is $\mathcal{H}$-orthogonal to each $\eta_j$,
\begin{equation}
\begin{aligned}
\langle\delta,\kappa_\perp(\cdot,x)\rangle_\mathcal{H}
&=\langle\delta,\kappa(\cdot,x)\rangle_\mathcal{H}{-}\sum_{j=1}^{n_\theta}\bigl(H^\dagger\eta(x)\bigr)_j\langle\delta,\eta_j\rangle_\mathcal{H}\\
&=\delta(x).
\end{aligned}
\label{eqn:ok.pf.repr}
\end{equation}
Hence, $\kappa_\perp$ satisfies both conditions of Definition~\ref{def:RKHS} on $\mathcal{H}_\perp$, and it is its reproducing kernel by uniqueness~\cite{aronszajn1950theory}.

\emph{Proof of (i).} Symmetry follows from the symmetry of $H^\dagger$. Continuity follows from the continuity of $\kappa$ and of $\eta_j\in\mathcal{H}$. By (iii), i.e., \eqref{eqn:ok.pf.repr}, applied to $\delta=\kappa_\perp(\cdot,x')\in\mathcal{H}_\perp$,
\[
\kappa_\perp(x,x')=\langle\kappa_\perp(\cdot,x'),\kappa_\perp(\cdot,x)\rangle_\mathcal{H};
\]
hence,
\[
\sum_{i,j}c_ic_j\kappa_\perp(x_i,x_j)=\Bigl\|\sum_ic_i\kappa_\perp(\cdot,x_i)\Bigr\|_\mathcal{H}^2\geq0,
\]
so that $\kappa_\perp$ is positive semidefinite. Similarly, with $v\doteq\sum_ic_i\eta(x_i)$,
\[
\sum_{i,j}c_ic_j\bigl(\kappa(x_i,x_j)-\kappa_\perp(x_i,x_j)\bigr)=v^\top H^\dagger v\geq0,
\]
since $H^\dagger\succeq0$.

\emph{Proof of (iv).} For $\delta\in\mathcal{H}$, $\Pi_\rho\delta=f^\top G^{-1}b(\delta)$ by \eqref{eqn:ok.projcf}. Since $G\succ0$, $\Pi_\rho\delta=0$ if and only if $b(\delta)=0$, i.e., if and only if $\delta\in\mathcal{H}_\perp$. The statement on kernel sections and their combinations follows from (ii), Step~2, and linearity.

\emph{Proof of (v).} Let $u\doteq\theta_1-\theta_2$, so that $f^\top u=\delta_2-\delta_1$. Taking physical moments of both sides, $b(f^\top u)=Gu$ on the left, and $b(\delta_2)-b(\delta_1)=0$ on the right, since $\delta_1,\delta_2\in\mathcal{H}_\perp$. Hence, $Gu=0$, so $u=0$ because $G\succ0$, and then $\delta_1=\delta_2$. \qed
\end{pf}
 
By Theorem~\ref{thm:ok.kperp}(v), within $\mathcal{H}_\perp$ the family \eqref{eqn:ok.ambiguity} reduces to a single member: no nonzero element of $\mathcal{F}$ is an admissible correction. This is a function-space statement; it does not exclude that some $\delta\in\mathcal{H}_\perp$ coincides with $f^\top a$ at finitely many sampling points.
 
Theorem~\ref{thm:ok.kperp} requires no condition on $H$; in particular, $H$ need not be invertible. By \eqref{eqn:ok.riesz}, $c^\top Hc=\|\eta^\top c\|^2_\mathcal{H}$ vanishes if and only if the physical direction $f^\top c$ is $\rho$-orthogonal to the whole of $\mathcal{H}$; in this case, the corresponding constraint in \eqref{eqn:ok.Hperp} is satisfied by every element of $\mathcal{H}$, and the pseudoinverse in \eqref{eqn:ok.kperp} accounts for this exactly.

Although the constraints in \eqref{eqn:ok.Hperp} are stated in the reference geometry, Theorem~\ref{thm:ok.kperp}(ii) obtains $\kappa_\perp$ through the projector $P_\perp$, which is orthogonal in the geometry of $\mathcal{H}$, and not through $\Pi_\rho$: $\langle\cdot,\cdot\rangle_\rho$ defines $\theta_\rho$, $\Delta_\perp$, and the constraints, whereas $\langle\cdot,\cdot\rangle_\mathcal{H}$ defines $P_\perp$ and the native geometry of $\kappa_\perp$. The map $P_\perp$ removes from $\delta\in\mathcal{H}$ a component in $\operatorname{span}\{\eta_1,\dots,\eta_{n_\theta}\}\subset\mathcal{H}$, whereas $\delta\mapsto\delta-\Pi_\rho\delta$ removes a component in $\mathcal{F}$, which in general does not belong to $\mathcal{H}$; both results are $\rho$-orthogonal to $\mathcal{F}$, but they differ in general. The consequences of using $\Pi_\rho$ are discussed in the following remark.
\begin{remark}[kernel projection by $\Pi_\rho$]\label{rem:ok.geometries}
Applying $\delta\mapsto\delta-\Pi_\rho\delta$ to the kernel sections in both arguments (in one argument only, the result is not symmetric in general) gives
\[
\begin{aligned}
\tilde\kappa(x,x')\doteq{}&\kappa(x,x')-f(x)^\top G^{-1}\eta(x')\\
&-\eta(x)^\top G^{-1}f(x')+f(x)^\top G^{-1}HG^{-1}f(x'),
\end{aligned}
\]
which is a positive definite kernel. Its native space is $\{\delta-\Pi_\rho\delta:\delta\in\mathcal{H}\}$, in general not a subspace of $\mathcal{H}$, and the norm of $\tilde\delta$ in it is the minimum of $\|\delta\|_\mathcal{H}$ over the $\delta\in\mathcal{H}$ such that $\delta-\Pi_\rho\delta=\tilde\delta$. Hence, by the change of variables $\theta\mapsto\theta-\vartheta_\rho(\delta)$, using $\tilde\kappa$ in Problem~\eqref{eqn:optprob.theta} is equivalent to solving it with $\kappa$ and reporting the re-labeled parameter $\vartheta_\rho(f^\top\theta^\star+\delta^\star)=\theta^\star+G^{-1}\sum_{j=1}^T\omega^\star_j\eta(x_j)$, i.e., the projected kernel calibration of~\cite{tuo2019adjustments} ($L^2_\rho$ calibration of the fitted map), which also targets $\theta_\rho$. This estimate re-labels the unconstrained fit, and its correction $\delta^\star-\Pi_\rho\delta^\star$ does not belong to $\mathcal{H}$ in general; $\kappa_\perp$ is instead the reproducing kernel of the subspace of $\mathcal{H}$ that satisfies the prescribed orthogonality, hence the term orthogonalized kernel.
\end{remark}

\subsection{Identification with the orthogonalized kernel}\label{sec:ok.ident}
 
Imposing the prescription of Section~\ref{sec:ok.canonical} on the correction of \eqref{eqn:optprob.theta} leads to the problem
\begin{equation}
(\theta^\star_\perp,\delta^\star_\perp)=\arg\!\!\min_{\theta,\,\delta\in\mathcal{H}_\perp}\sum_{t=1}^{T}\bigl[y_t-\hat y_t(\theta,\delta)\bigr]^2+\gamma\|\delta\|^2_\mathcal{H},
\label{eqn:ok.optprob}
\end{equation}
where $\hat y_t(\theta,\delta)$ is as in \eqref{eqn:optprob.theta} and $\theta$ ranges over $\mathbb{R}^{n_\theta}$, that is, Problem~\eqref{eqn:optprob.theta} subject to the constraint $\Pi_\rho\delta=0$ (see Theorem~\ref{thm:ok.kperp}(iv)). Consistently with the notation of \eqref{eqn:Psidef}--\eqref{eqn:thmlincf}, let
\begin{subequations}
\begin{align}
E(x)&\doteq[\eta(x_1),\dots,\eta(x_T)]^\top\in\mathbb{R}^{T\times n_\theta},\\
\mathbf{K}_\perp&\doteq\mathbf{K}-E(x)\,H^\dagger E(x)^\top,\\
\Psi_\perp&\doteq(\mathbf{K}_\perp+\gamma\mathbf{I}_T)^{-1},
\end{align}
\label{eqn:ok.Kperp}%
\end{subequations}
so that $(\mathbf{K}_\perp)_{ij}=\kappa_\perp(x_i,x_j)$.
 
\begin{thm}[orthogonal model embedding]\label{thm:ok.embedding}
Let Assumptions~\ref{ass:ok.geometry} and~\ref{ass:ok.kernel} hold, let $\gamma>0$, and let $F(x)$ have full column rank. Then, Problem~\eqref{eqn:ok.optprob} has the unique solution
\begin{subequations}
\begin{align}
\theta^\star_\perp&=\bigl(F(x)^\top\Psi_\perp F(x)\bigr)^{-1}F(x)^\top\Psi_\perp Y,
\label{eqn:ok.theta}\\
\delta^\star_\perp(x)&=\textstyle\sum_{j=1}^{T}\omega^\star_{\perp,j}\,\kappa_\perp(x,x_j),
\label{eqn:ok.delta}
\end{align}
\label{eqn:ok.solution}%
\end{subequations}
with $\omega^\star_\perp\doteq\Psi_\perp\bigl(Y-F(x)\theta^\star_\perp\bigr)$. Moreover, $\Pi_\rho\delta^\star_\perp=0$ and, defining the fitted map $\hat g_\perp\doteq f^\top\theta^\star_\perp+\delta^\star_\perp$,
\begin{equation}
\theta^\star_\perp=\vartheta_\rho(\hat g_\perp).
\label{eqn:ok.selfconsistent}
\end{equation}
\end{thm}
 
\begin{pf}
By Theorem~\ref{thm:ok.kperp}(i) and (iii), $\kappa_\perp$ satisfies Definition~\ref{def:pdkernel} and $(\mathcal{H}_\perp,\|\cdot\|_\mathcal{H})$ is its RKHS. Hence, \eqref{eqn:ok.optprob} is Problem~\eqref{eqn:optprob.theta} with $(\mathcal{H},\kappa,\mathbf{K})$ replaced by $(\mathcal{H}_\perp,\kappa_\perp,\mathbf{K}_\perp)$, and \eqref{eqn:ok.solution} follows from \eqref{eqn:reprthm.ext} and \eqref{eqn:thmlincf}. Since $\mathbf{K}_\perp\succeq0$, $\Psi_\perp\succ0$ and, as $F(x)$ has full column rank, $F(x)^\top\Psi_\perp F(x)\succ0$; thus, the reduced cost in $\theta$ is strictly convex, and, for $\gamma>0$, the inner kernel ridge regression problem has a unique minimizer. Finally, $\delta^\star_\perp\in\mathcal{H}_\perp$, so $\Pi_\rho\delta^\star_\perp=0$ by Theorem~\ref{thm:ok.kperp}(iv), and \eqref{eqn:ok.selfconsistent} follows from linearity of $\vartheta_\rho$ and $\vartheta_\rho(f^\top\theta^\star_\perp)=\theta^\star_\perp$. \qed
\end{pf}

Theorem~\ref{thm:ok.embedding} leaves the computational structure of \eqref{eqn:thmlincf} unchanged, the only modification being $\mathbf{K}\to\mathbf{K}_\perp$. Since $\kappa-\kappa_\perp$ is positive semidefinite, $\mathbf{K}_\perp\preceq\mathbf{K}$, hence $\Psi_\perp\succeq\Psi$ and $F(x)^\top\Psi_\perp F(x)\succeq F(x)^\top\Psi F(x)$: the curvature of the reduced cost in $\theta$ is not decreased along any direction. By \eqref{eqn:ok.selfconsistent} and Definition~\ref{def:ok.canonical}, the estimate and its target are obtained by applying the same linear map $\vartheta_\rho$ to $\hat g_\perp$ and to $\bar g$, respectively; hence, the error on the physical parameter is controlled by the error with which $\hat g_\perp$ reconstructs $\bar g$ on $\Omega$, as stated next.
 
\begin{corollary}[parameter error bound]\label{cor:ok.bound}
Under the assumptions of Theorem~\ref{thm:ok.embedding} and Definition~\ref{def:ok.canonical},
\begin{equation}
\|\theta^\star_\perp-\theta_\rho\|_2\leq\lambda_{\min}(G)^{-1/2}\,\|\hat g_\perp-\bar g\|_\rho.
\label{eqn:ok.bound}
\end{equation}
\end{corollary}
 
\begin{pf}
By \eqref{eqn:ok.selfconsistent}, Definition~\ref{def:ok.canonical}, and linearity, $\theta^\star_\perp-\theta_\rho=\vartheta_\rho(\hat g_\perp-\bar g)$, where $\hat g_\perp-\bar g\in L^2_\rho$ by Assumptions~\ref{ass:ok.geometry} and~\ref{ass:ok.kernel}. The claim follows from Proposition~\ref{prop:ok.proj}(iv). \qed
\end{pf}
 
The bound \eqref{eqn:ok.bound} is deterministic and holds for every data set; consistency would further require conditions on the input design, the noise, and the regularization ensuring that $\|\hat g_\perp-\bar g\|_\rho\to0$. The bound holds for any estimator of the form $\vartheta_\rho(\hat g)$, including the re-labeled unconstrained fit of Remark~\ref{rem:ok.geometries}; the orthogonalized kernel differs from it in that the constraint restricts the hypothesis class, and hence modifies the fitted map itself.

Since the constraint $\Pi_\rho\delta=0$ is imposed on functions rather than on samples, \eqref{eqn:ok.Kperp}--\eqref{eqn:ok.solution} is exact for \eqref{eqn:ok.optprob}, and $\delta^\star_\perp$ is exactly $\rho$-orthogonal to $\mathcal{F}$. The only integrals involved, $\eta$ and $H$ in \eqref{eqn:ok.eta}--\eqref{eqn:ok.H}, depend on $(f,\Omega,\rho,\kappa)$ but not on the data, and are computed before, and independently of, the identification problem. 

\section{Set-membership identifiability}\label{sec:sm}

While Corollary~\ref{cor:ok.bound} bounds the error on $\theta_\rho$ by the error with which the fitted map reconstructs $\bar g$ on $\Omega$, it does not describe which parameters are compatible with the data, nor how accurately $\theta_\rho$ can be recovered from them. This section addresses both questions in the set-membership framework~\cite{milanese1991optimal,milanese2004set}, in which the noise and the discrepancy are bounded, and compares the correction classes $\mathcal{H}$ and $\mathcal{H}_\perp$ independently of the estimator. For purely nonparametric kernel models, in which the unknown function lies in a ball of an RKHS and the noise is bounded pointwise, deterministic error bounds for kernel ridge regression and support vector regression were derived in~\cite{maddalena2021deterministic}, and the tightest bounds on the value of the function at a given point were computed by finite-dimensional convex programs in~\cite{scharnhorst2023robust}.

\subsection{Feasible parameter sets}\label{sec:sm.fps}

For a function $g$ on $\mathcal{X}$, let $[g]_\mathcal{D}\doteq[g(x_1),\dots,g(x_T)]^\top$ denote its values at the inputs of the data set $\mathcal{D}$.

\begin{assumption}[set-membership information]\label{ass:sm.info}
The restriction of $\Delta$ to $\Omega$ belongs to $L^2_\rho$, and: (i) $Y-F(x)\theta_\rho-[\Delta_\perp]_\mathcal{D}\in\mathcal{E}$, where $\mathcal{E}\subset\mathbb{R}^T$ is a known compact, convex set with $\mathcal{E}=-\mathcal{E}$; (ii) $\Delta_\perp\in\mathcal{H}$ and $\|\Delta_\perp\|_\mathcal{H}\leq\beta$, for a known $\beta>0$.
\end{assumption}

By Proposition~\ref{prop:ok.decomp}(ii), the vector in item \textit{(i)} collects the noise samples $e$ of \eqref{eqn:system.comp.lin}. We denote by $\nu(z)\doteq\max_{e\in\mathcal{E}}z^\top e$, $z\in\mathbb{R}^T$, the support function of $\mathcal{E}$. Two common choices are the box $\mathcal{E}=\{e:\|e\|_\infty\leq\varepsilon\}$, for which $\nu(z)=\varepsilon\|z\|_1$, and the energy bound $\mathcal{E}=\{e:\|e\|_2\leq\mu\}$~\cite{fogel1979system}, for which $\nu(z)=\mu\|z\|_2$. Since $\Delta_\perp\perp_\rho\mathcal{F}$ by Proposition~\ref{prop:ok.decomp}, item \textit{(ii)} implies $\Delta_\perp\in\mathcal{H}_\perp$; the bound $\beta$ thus concerns the component of the discrepancy that the physical model cannot reproduce on $\Omega$. 
In general, item \textit{(ii)} neither implies nor is implied by the assumption $\Delta\in\mathcal{H}$ of~\cite{donati2025kernel}, since $\Delta-\Delta_\perp=f^\top\theta_\Delta$. If $\mathcal{F}\subset\mathcal{H}$, the two assumptions are equivalent. If, instead, no nonzero element of $\mathcal{F}$ belongs to $\mathcal{H}$, they hold simultaneously only if $\theta_\Delta=0$, i.e., if $\Delta$ is already $\rho$-orthogonal to $\mathcal{F}$: indeed, both would give $f^\top\theta_\Delta\in\mathcal{F}\cap\mathcal{H}=\{0\}$, hence $\theta_\Delta=0$, since $G\succ0$.

The admissible error sets of the correction classes $\mathcal{H}$ and $\mathcal{H}_\perp$ are
\begin{equation}
\begin{aligned}
\mathcal{W}&\doteq\bigl\{[\delta]_\mathcal{D}+e:\ \delta\in\mathcal{H},\ \|\delta\|_\mathcal{H}\leq\beta,\ e\in\mathcal{E}\bigr\},\\
\mathcal{W}_\perp&\doteq\bigl\{[\delta]_\mathcal{D}+e:\ \delta\in\mathcal{H}_\perp,\ \|\delta\|_\mathcal{H}\leq\beta,\ e\in\mathcal{E}\bigr\}.
\end{aligned}
\label{eqn:sm.W}
\end{equation}

\begin{definition}[feasible parameter sets]\label{def:sm.fps}
Under Assumptions~\ref{ass:ok.geometry} and~\ref{ass:ok.kernel}, the feasible parameter sets associated with Problems~\eqref{eqn:optprob.theta} and~\eqref{eqn:ok.optprob} are, respectively,
\begin{align*}
\mathcal{S}(\mathcal{D})&\doteq\{\theta\in\mathbb{R}^{n_\theta}:\ Y-F(x)\theta\in\mathcal{W}\},\\
\mathcal{S}_\perp(\mathcal{D})&\doteq\{\theta\in\mathbb{R}^{n_\theta}:\ Y-F(x)\theta\in\mathcal{W}_\perp\}.
\end{align*}
\end{definition}

Thus, $\mathcal{S}(\mathcal{D})$ contains the parameters for which some $\delta\in\mathcal{H}$ with $\|\delta\|_\mathcal{H}\leq\beta$ reproduces the data up to an admissible noise, and $\mathcal{S}_\perp(\mathcal{D})$ is the analogous set with $\delta\in\mathcal{H}_\perp$.

\begin{proposition}[admissible error sets]\label{prop:sm.W}
Under Assumptions~\ref{ass:ok.geometry} and~\ref{ass:ok.kernel}:
\begin{itemize}
\item[(i)] $\mathcal{W}_\perp=\{\mathbf{K}_\perp\omega+e:\ \omega\in\mathbb{R}^T,\ \omega^\top\mathbf{K}_\perp\omega\leq\beta^2,\ e\in\mathcal{E}\}$, and $\mathcal{W}$ is obtained by replacing $\mathbf{K}_\perp$ with $\mathbf{K}$;
\item[(ii)] $\mathcal{W}_\perp\subseteq\mathcal{W}$, and both sets are compact, convex, and symmetric;
\item[(iii)] for every $z\in\mathbb{R}^T$,
\begin{equation}
\begin{aligned}
\max_{w\in\mathcal{W}_\perp}z^\top w&=\varphi_\perp(z)\doteq\beta\,\|\mathbf{K}_\perp^{1/2}z\|_2+\nu(z),\\
\max_{w\in\mathcal{W}}z^\top w&=\varphi(z)\doteq\beta\,\|\mathbf{K}^{1/2}z\|_2+\nu(z),
\end{aligned}
\label{eqn:sm.phi}
\end{equation}
and $\varphi_\perp(z)\leq\varphi(z)$.
\end{itemize}
\end{proposition}

\begin{pf}
\textit{(i)} Let $\delta\in\mathcal{H}_\perp$ with $\|\delta\|_\mathcal{H}\leq\beta$. The functions $\kappa_\perp(\cdot,x_t)$, $t=1,\dots,T$, belong to $\mathcal{H}_\perp$ by Theorem~\ref{thm:ok.kperp}(iii); let $\tilde\delta$ be the $\langle\cdot,\cdot\rangle_\mathcal{H}$-orthogonal projection of $\delta$ onto their span. Since $\delta-\tilde\delta$ is $\mathcal{H}$-orthogonal to each $\kappa_\perp(\cdot,x_t)$, \eqref{eqn:ok.pf.repr} gives
$$
\begin{aligned}
\delta(x_t)&=\langle\delta,\kappa_\perp(\cdot,x_t)\rangle_\mathcal H\\&=\langle\tilde\delta,\kappa_\perp(\cdot,x_t)\rangle_\mathcal H+\underbrace{\langle\delta-\tilde\delta,\kappa_\perp(\cdot,x_t)\rangle_\mathcal H}_{=0}=\tilde\delta(x_t)
\end{aligned}
$$
and $\|\tilde\delta\|_\mathcal{H}\leq\|\delta\|_\mathcal{H}\leq\beta$. Writing $\tilde\delta=\sum_j\omega_j\kappa_\perp(\cdot,x_j)$ and applying \eqref{eqn:ok.pf.repr} with $\kappa_\perp(\cdot,x_j)$ in place of $\delta$, i.e., $\langle\kappa_\perp(\cdot,x_j),\kappa_\perp(\cdot,x_t)\rangle_\mathcal{H}=\kappa_\perp(x_t,x_j)$, gives $[\tilde\delta]_\mathcal{D}=\mathbf{K}_\perp\omega$ and $\|\tilde\delta\|_\mathcal{H}^2=\omega^\top\mathbf{K}_\perp\omega$. Hence, $[\delta]_\mathcal{D}=\mathbf{K}_\perp\omega$ with $\omega^\top\mathbf{K}_\perp\omega\leq\beta^2$. Conversely, for every $\omega$ with $\omega^\top\mathbf{K}_\perp\omega\leq\beta^2$, the function $\sum_j\omega_j\kappa_\perp(\cdot,x_j)$ belongs to $\mathcal{H}_\perp$, has norm at most $\beta$, and has samples $\mathbf{K}_\perp\omega$. Adding $e\in\mathcal{E}$ proves the claim for $\mathcal{W}_\perp$. The same argument applies to $\mathcal{W}$, with $\kappa$, $\mathcal{H}$, and the reproducing property of Definition~\ref{def:RKHS} in place of $\kappa_\perp$, $\mathcal{H}_\perp$, and \eqref{eqn:ok.pf.repr}.

\textit{(ii)} The inclusion follows from $\mathcal{H}_\perp\subseteq\mathcal{H}$. By (i), $\mathcal{W}_\perp$ is the sum of $\mathcal{E}$ and $\{\mathbf{K}_\perp\omega:\omega^\top\mathbf{K}_\perp\omega\leq\beta^2\}=\{\mathbf{K}_\perp^{1/2}s:\|s\|_2\leq\beta\}$, where the equality follows by setting $s=\mathbf{K}_\perp^{1/2}\omega$ and, conversely, $\omega=(\mathbf{K}_\perp^{1/2})^\dagger s$. Both terms are compact, convex, and symmetric, and so is their sum.

\textit{(iii)} The maximum of $z^\top w$ over a sum of sets is the sum of the maxima, and $\max_{\|s\|_2\leq\beta}z^\top\mathbf{K}_\perp^{1/2}s=\beta\|\mathbf{K}_\perp^{1/2}z\|_2$. Finally, $\mathbf{K}_\perp\preceq\mathbf{K}$ by Theorem~\ref{thm:ok.kperp}(i), so that $\|\mathbf{K}_\perp^{1/2}z\|_2\leq\|\mathbf{K}^{1/2}z\|_2$. \qed
\end{pf}

\begin{proposition}[properties of the feasible sets]\label{prop:sm.fps}
Let Assumptions~\ref{ass:ok.geometry}, \ref{ass:ok.kernel}, and~\ref{ass:sm.info} hold. Then: (i) $\theta_\rho\in\mathcal{S}_\perp(\mathcal{D})\subseteq\mathcal{S}(\mathcal{D})$; (ii) both sets are convex and, if $F(x)$ has full column rank, compact.
\end{proposition}

\begin{pf}
\textit{(i)} By Assumption~\ref{ass:sm.info}, $Y-F(x)\theta_\rho=[\Delta_\perp]_\mathcal{D}+e$ with $\Delta_\perp\in\mathcal{H}_\perp$, $\|\Delta_\perp\|_\mathcal{H}\leq\beta$, and $e\in\mathcal{E}$, i.e., it belongs to $\mathcal{W}_\perp$. The inclusion follows from $\mathcal{W}_\perp\subseteq\mathcal{W}$. \textit{(ii)} Both sets are preimages of compact convex sets under the affine map $\theta\mapsto Y-F(x)\theta$, hence closed and convex; if $F(x)$ has full column rank, they are bounded, since every $\theta$ in either set satisfies $\lambda_{\min}(F(x)^\top F(x))^{1/2}\|\theta\|_2\leq\|F(x)\theta\|_2\leq\|Y\|_2+\max_{w\in\mathcal{W}}\|w\|_2$, where the maximum is finite by Proposition~\ref{prop:sm.W}(ii) and $\lambda_{\min}(F(x)^\top F(x))>0$. \qed
\end{pf}

Every $\theta\in\mathcal{S}_\perp(\mathcal{D})$ is the canonical parameter of a map $f^\top\theta+\delta$, $\delta\in\mathcal{H}_\perp$, that satisfies Assumption~\ref{ass:sm.info} for the same data. Hence, the data and Assumption~\ref{ass:sm.info} cannot distinguish $\theta_\rho$ from any element of $\mathcal{S}_\perp(\mathcal{D})$, and $\mathcal{S}_\perp(\mathcal{D})$ is the smallest set that is guaranteed to contain $\theta_\rho$ on the basis of this information.
By Proposition~\ref{prop:sm.W}(i), $\theta\in\mathcal{S}_\perp(\mathcal{D})$ if and only if there exists $\omega\in\mathbb{R}^T$ such that
\[
\|\mathbf{K}_\perp^{1/2}\omega\|_2\leq\beta,\qquad Y-F(x)\theta-\mathbf{K}_\perp\omega\in\mathcal{E}.
\]
Hence, the smallest and the largest values of $\theta_j$ over $\mathcal{S}_\perp(\mathcal{D})$ are the optimal values of the minimization and the maximization of $\theta_j$ over $(\theta,\omega)$ subject to these constraints. These are convex programs in $n_\theta+T$ variables; the second constraint consists of $2T$ linear inequalities for the box bound and of one second-order cone constraint for the energy bound, so that both are second-order cone programs. Since $\mathcal{S}_\perp(\mathcal{D})$ is convex and, if $F(x)$ has full column rank, compact, the $2n_\theta$ optimal values give the tightest interval containing each component of $\theta$. The same holds for $\mathcal{S}(\mathcal{D})$, with $\mathbf{K}$ in place of $\mathbf{K}_\perp$.

\subsection{Best achievable accuracy}\label{sec:sm.acc}
The feasible set $\mathcal{S}_\perp(\mathcal{D})$ collects the parameters that cannot be distinguished from $\theta_\rho$. The accuracy that an estimator can guarantee on $\theta_\rho$ is quantified next. An estimator is a map $\chi:\mathbb{R}^T\to\mathbb{R}^{n_\theta}$ that assigns an estimate $\chi(Y)$ to the data vector $Y$, the inputs $x_1,\dots,x_T$ being fixed. By the discussion after Proposition~\ref{prop:sm.fps}, every pair $\theta\in\mathbb{R}^{n_\theta}$, $w\in\mathcal{W}_\perp$ corresponds to a map that satisfies Assumption~\ref{ass:sm.info}, has canonical parameter $\theta$, and produces the data $Y=F(x)\theta+w$. Hence, the worst-case error of $\chi$ on the $j$-th entry of $\theta_\rho$ is
\[
\operatorname{err}_j(\chi)\doteq\sup\bigl\{|\chi_j(F(x)\theta+w)-\theta_j|:\theta\in\mathbb{R}^{n_\theta},\,w\in\mathcal{W}_\perp\bigr\},
\]
where $\chi_j$ denotes the $j$-th entry of $\chi$. Let $\iota_j$ denote the $j$-th column of $\mathbf{I}_{n_\theta}$, and
\[
\operatorname{err}^\star_j\doteq\max\bigl\{\theta_j:\ F(x)\theta\in\mathcal{W}_\perp\bigr\},\qquad j=1,\dots,n_\theta,
\]
i.e., the extent along $\theta_j$ of the feasible set $\mathcal{S}_\perp(\mathcal{D})$ obtained for $Y=0$.

\begin{proposition}[best achievable accuracy]\label{prop:sm.acc}
Let Assumptions~\ref{ass:ok.geometry} and~\ref{ass:ok.kernel} hold, let $F(x)$ have full column rank, and let either $\mathbf{K}_\perp\succ0$ or $\mathcal{E}$ contain a neighborhood of the origin. Then, for $j=1,\dots,n_\theta$:
\begin{itemize}
\item[(i)] every estimator $\chi$ satisfies $\operatorname{err}_j(\chi)\geq\operatorname{err}^\star_j$;
\item[(ii)] every linear estimator $\chi(Y)=Z^\top Y$, with $Z\in\mathbb{R}^{T\times n_\theta}$ such that $F(x)^\top Z=\mathbf{I}_{n_\theta}$, satisfies $\operatorname{err}_j(\chi)=\varphi_\perp(z_j)$, where $z_j$ is the $j$-th column of $Z$;
\item[(iii)] $\operatorname{err}^\star_j=\min\{\varphi_\perp(z):\ z\in\mathbb{R}^T,\ F(x)^\top z=\iota_j\}$, and the minimum is attained;
\item[(iv)] if $\mathbf{K}_\perp\succ0$ and $\mathcal{E}=\{e:\|e\|_2\leq\mu\}$ with $\mu\geq0$, then there exists $\gamma_j\geq0$, with $\gamma_j>0$ if $\mu>0$, such that the $j$-th entry of $\theta^\star_\perp$ in~\eqref{eqn:ok.theta}, computed with $\gamma=\gamma_j$, has worst-case error $\operatorname{err}^\star_j$.
\end{itemize}
\end{proposition}
\begin{pf}
\textit{(i)} The set $\{\theta:F(x)\theta\in\mathcal{W}_\perp\}$ contains $\theta=0$, since $0\in\mathcal{W}_\perp$, is closed, since $\mathcal{W}_\perp$ is compact by Proposition~\ref{prop:sm.W}(ii), and is bounded, since
\[
\lambda_{\min}(F(x)^\top F(x))^{1/2}\|\theta\|_2\leq\|F(x)\theta\|_2\leq\max_{w\in\mathcal{W}_\perp}\|w\|_2,
\]
with $\lambda_{\min}(F(x)^\top F(x))>0$. Hence, the maximum defining $\operatorname{err}^\star_j$ is attained at some $\theta^0$, and $\theta^0_j=\operatorname{err}^\star_j\geq0$. Since $\mathcal{W}_\perp=-\mathcal{W}_\perp$, the data $Y=0$ are produced both by $\theta=\theta^0$ with $w=-F(x)\theta^0\in\mathcal{W}_\perp$ and by $\theta=-\theta^0$ with $w=F(x)\theta^0\in\mathcal{W}_\perp$. Every estimator returns the same $\chi(0)$ in the two cases, and
\[
|\chi_j(0)-\theta^0_j|+|\chi_j(0)+\theta^0_j|\geq2\theta^0_j=2\operatorname{err}^\star_j;
\]
hence, one of the two errors is at least $\operatorname{err}^\star_j$.

\textit{(ii)} For $Y=F(x)\theta+w$, $F(x)^\top z_j=\iota_j$ gives $z_j^\top Y-\theta_j=z_j^\top w$. Since $\mathcal{W}_\perp$ is symmetric, $\sup_{w\in\mathcal{W}_\perp}|z_j^\top w|=\max_{w\in\mathcal{W}_\perp}z_j^\top w=\varphi_\perp(z_j)$, by Proposition~\ref{prop:sm.W}(iii).

\textit{(iii)} Let $z\in\mathbb{R}^T$ satisfy $F(x)^\top z=\iota_j$. If $F(x)\theta\in\mathcal{W}_\perp$, then $\theta_j=z^\top F(x)\theta\leq\varphi_\perp(z)$ by Proposition~\ref{prop:sm.W}(iii); hence, $\operatorname{err}^\star_j\leq\varphi_\perp(z)$. To show that equality holds for some $z$, note that $\mathcal{W}_\perp$ contains a neighborhood of the origin: if $\mathbf{K}_\perp\succ0$, it contains $\{\mathbf{K}_\perp^{1/2}s:\|s\|_2\leq\beta\}$ (see the proof of Proposition~\ref{prop:sm.W}(ii)), and otherwise it contains $\mathcal{E}$. Hence, $-\operatorname{err}^\star_j$ is the optimal value of the convex problem of minimizing $-\theta_j$ over $(\theta,w)\in\mathbb{R}^{n_\theta}\times\mathcal{W}_\perp$ subject to $F(x)\theta=w$, for which the feasible point $(0,0)$ lies in the interior of the domain. By Slater's condition, strong duality holds and, since $\operatorname{err}^\star_j$ is finite by (i), the dual optimum is attained~\cite[Sec.~5.2.3]{boyd2004convex}. With multiplier $z\in\mathbb{R}^T$ for the constraint $F(x)\theta=w$, the dual function is
\[
\begin{aligned}
&\inf_{\theta,w\in\mathcal{W}_\perp}\bigl\{-\theta_j+z^\top(F(x)\theta-w)\bigr\}\\
&\qquad=\begin{cases}-\varphi_\perp(z),&F(x)^\top z=\iota_j,\\ -\infty,&\text{otherwise},\end{cases}
\end{aligned}
\]
so that $\operatorname{err}^\star_j=\min\{\varphi_\perp(z):F(x)^\top z=\iota_j\}$.

\textit{(iv)} Let $\hat z_j$ attain the minimum in (iii). Since $F(x)^\top\hat z_j=\iota_j$, $\hat z_j\neq0$, and, since $\mathbf{K}_\perp\succ0$, $\mathbf{K}_\perp^{1/2}\hat z_j\neq0$. Hence, $\varphi_\perp(z)=\beta\|\mathbf{K}_\perp^{1/2}z\|_2+\mu\|z\|_2$ is differentiable at $\hat z_j$. For every $v$ with $F(x)^\top v=0$, the map $t\mapsto\varphi_\perp(\hat z_j+tv)$ is minimized at $t=0$, so that $\nabla\varphi_\perp(\hat z_j)^\top v=0$; hence, $\nabla\varphi_\perp(\hat z_j)$ lies in the range of $F(x)$, i.e., there exists $\lambda\in\mathbb{R}^{n_\theta}$ such that
\[
\beta\,\frac{\mathbf{K}_\perp\hat z_j}{\|\mathbf{K}_\perp^{1/2}\hat z_j\|_2}+\mu\,\frac{\hat z_j}{\|\hat z_j\|_2}=F(x)\lambda.
\]
Multiplying by $\|\mathbf{K}_\perp^{1/2}\hat z_j\|_2/\beta$ gives $(\mathbf{K}_\perp+\gamma_j\mathbf{I}_T)\hat z_j=F(x)c$, with
\[
\gamma_j\doteq\frac{\mu\,\|\mathbf{K}_\perp^{1/2}\hat z_j\|_2}{\beta\,\|\hat z_j\|_2},
\qquad
c\doteq\frac{\|\mathbf{K}_\perp^{1/2}\hat z_j\|_2}{\beta}\,\lambda,
\]
where $\gamma_j\geq0$, and $\gamma_j>0$ if $\mu>0$. Let $\Psi_\perp=(\mathbf{K}_\perp+\gamma_j\mathbf{I}_T)^{-1}$, which exists since $\mathbf{K}_\perp\succ0$. Then, $\hat z_j=\Psi_\perp F(x)c$, and $F(x)^\top\hat z_j=\iota_j$ gives \[ c=\bigl(F(x)^\top\Psi_\perp F(x)\bigr)^{-1}\iota_j, \] where $F(x)^\top\Psi_\perp F(x)\succ0$ since $F(x)$ has full column rank. Therefore, by~\eqref{eqn:ok.theta} and the symmetry of $\Psi_\perp$, the $j$-th entry of $\theta^\star_\perp$ computed with $\gamma=\gamma_j$ is $\hat z_j^\top Y$, and its worst-case error is $\varphi_\perp(\hat z_j)=\operatorname{err}^\star_j$, by (ii) and (iii). \qed
\end{pf}

By Proposition~\ref{prop:sm.acc}(i)--(iii), $\operatorname{err}^\star_j$ is the best accuracy that any estimator can guarantee on the $j$-th entry of $\theta_\rho$. It depends on the inputs, on $\mathbf{K}_\perp$, $\beta$, and $\mathcal{E}$, but not on the outputs, and it is computed by $n_\theta$ second-order cone programs. It also bounds the size of the feasible set: if $\theta,\theta'\in\mathcal{S}_\perp(\mathcal{D})$, then $F(x)(\theta-\theta')/2\in\mathcal{W}_\perp$ by convexity and symmetry, so that $|\theta_j-\theta'_j|\leq2\operatorname{err}^\star_j$. Item~(ii) applies to all the linear estimators considered so far: the KME estimate~\eqref{eqn:thmlincf}, with $Z=\Psi F(x)\bigl(F(x)^\top\Psi F(x)\bigr)^{-1}$, the least-squares estimate, with $Z=F(x)\bigl(F(x)^\top F(x)\bigr)^{-1}$, and the re-labeled estimate of Remark~\ref{rem:ok.geometries}, which is linear in $Y$ and equals $\theta$ when $Y=F(x)\theta$. Hence, their worst-case errors on $\theta_\rho$, under the same information, are given in closed form by $\varphi_\perp$ and can be compared with $\operatorname{err}^\star_j$. Item~(iv) shows that, with energy-bounded noise, this optimum is attained by the orthogonalized estimator itself, i.e., by the regularized estimate in the class $\mathcal{F}\oplus\mathcal{H}_\perp$, for a suitable weight $\gamma_j$, in line with classical results of optimal recovery from inaccurate data~\cite{micchelli1977survey,melkman1979optimal}. For function values, the tightest bounds under the same noise model are expressed through the posterior mean and variance of a Gaussian process with an optimally chosen noise variance~\cite{lahr2025optimal}. Moreover, if $\mu>0$, the minimizer in (iii), denoted by $\hat z_j$, is unique, since $\|z\|_2$ is strictly convex on any affine set that does not contain the origin; hence, every linear estimator whose $j$-th column differs from $\hat z_j$ has a worst-case error strictly larger than $\operatorname{err}^\star_j$. The weight $\gamma_j$ depends on $j$, $\beta$, and $\mu$; hence, item~(iv) characterizes the optimal estimator rather than providing a tuning rule. Noise sets that are not Euclidean balls, such as the box, are not covered by item~(iv).

\subsection{Identifiability with and without orthogonality}\label{sec:sm.ident}
The next result shows that $\mathcal{S}_\perp(\mathcal{D})$ shrinks to $\theta_\rho$ as the data become dense in $\Omega$ and the noise vanishes. Consider a sequence of data sets $\mathcal{D}_k=\{(x_{k,t},y_{k,t})\}_{t=1}^{T_k}$, $k\in\mathbb{N}$, with $x_{k,t}\in\Omega$, and let
\[
\tau_k\doteq\sup_{x\in\Omega}\ \min_{t=1,\dots,T_k}\|x-x_{k,t}\|_2
\]
denote the fill distance of the inputs of $\mathcal{D}_k$ in $\Omega$.

\begin{thm}[convergence of the feasible set]\label{thm:sm.conv}
Let Assumptions~\ref{ass:ok.geometry} and~\ref{ass:ok.kernel} hold, and let each $\mathcal{D}_k$ satisfy Assumption~\ref{ass:sm.info} with the same $\beta$ and with a noise set $\mathcal{E}_k\subseteq\{e:\|e\|_\infty\leq\varepsilon_k\}$. If $\varepsilon_k\to0$ and $\tau_k\to0$, then
\[
\lim_{k\to\infty}\ \sup_{\theta\in\mathcal{S}_\perp(\mathcal{D}_k)}\|\theta-\theta_\rho\|_2=0.
\]
\end{thm}
\begin{pf}
Assume, by contradiction, that there exist $\alpha>0$ and, after passing to a subsequence, $\theta_k\in\mathcal{S}_\perp(\mathcal{D}_k)$ such that $\|\theta_k-\theta_\rho\|_2\geq\alpha$ for all $k$. With $Y$ and $F(x)$ formed from $\mathcal{D}_k$, Definition~\ref{def:sm.fps} and Assumption~\ref{ass:sm.info} give $\delta_k\in\mathcal{H}_\perp$, with $\|\delta_k\|_\mathcal{H}\leq\beta$, and $e_k,e'_k\in\mathcal{E}_k$ such that $Y-F(x)\theta_k=[\delta_k]_{\mathcal{D}_k}+e_k$ and $Y-F(x)\theta_\rho=[\Delta_\perp]_{\mathcal{D}_k}+e'_k$. Let $a_k\doteq\theta_k-\theta_\rho$ and $d_k\doteq\delta_k-\Delta_\perp$, which belongs to $\mathcal{H}_\perp$, since $\Delta_\perp\in\mathcal{H}_\perp$, and satisfies $\|d_k\|_\mathcal{H}\leq2\beta$. Subtracting the two identities gives $|f(x_{k,t})^\top a_k+d_k(x_{k,t})|\leq2\varepsilon_k$, $t=1,\dots,T_k$.

Let $c_k\doteq\max\{1,\|a_k\|_2\}$, $b_k\doteq a_k/c_k$, and $h_k\doteq d_k/c_k\in\mathcal{H}_\perp$. Then, $\min\{1,\alpha\}\leq\|b_k\|_2\leq1$, $\|h_k\|_\mathcal{H}\leq2\beta$, and $u_k\doteq f^\top b_k+h_k$ satisfies $|u_k(x_{k,t})|\leq2\varepsilon_k$, $t=1,\dots,T_k$. By \eqref{eqn:ok.pointbound}, $|h_k(x)|\leq2\beta\,\kappa(x,x)^{1/2}$; moreover, $|h_k(x)-h_k(x')|\leq2\beta\|\kappa(\cdot,x)-\kappa(\cdot,x')\|_\mathcal{H}$, where $\|\kappa(\cdot,x)-\kappa(\cdot,x')\|_\mathcal{H}^2=\kappa(x,x)-2\kappa(x,x')+\kappa(x',x')$ tends to zero uniformly as $\|x-x'\|_2\to0$, since $\kappa$ is uniformly continuous on the compact set $\Omega\times\Omega$. Hence, the continuous functions $h_k$ are pointwise bounded and equicontinuous on $\Omega$, and, by the Arzel\`a--Ascoli theorem~\cite[Definitions~7.19 and~7.22, Theorem~7.25]{rudin1976principles}, after passing to a further subsequence, $h_k\to h$ uniformly on $\Omega$, with $h$ continuous, and $b_k\to b$, with $\|b\|_2\geq\min\{1,\alpha\}$. Since $f$ is bounded on $\Omega$, the uniform convergence gives $\langle f_j,h\rangle_\rho=\lim_k\langle f_j,h_k\rangle_\rho=0$, $j=1,\dots,n_\theta$, and $u_k\to u\doteq f^\top b+h$ uniformly on $\Omega$.

Fix $x\in\Omega$ at which $f$, and hence $u$, is continuous, and choose, for each $k$, an index $t_k$ such that $\|x-x_{k,t_k}\|_2\leq\tau_k$. Then,
\[
|u(x)|\leq|u(x)-u(x_{k,t_k})|+\sup_{\xi\in\Omega}|u(\xi)-u_k(\xi)|+|u_k(x_{k,t_k})|,
\]
where the first term tends to zero by the continuity of $u$ at $x$, since $\tau_k\to0$, the second by the uniform convergence, and the third since it is at most $2\varepsilon_k$. Hence, $u$ vanishes at every point of continuity of $f$, i.e., on $\Omega$ under Assumption~\ref{ass:ok.geometry}(iii), and almost everywhere on $\Omega$ if $f$ is only piecewise continuous. Multiplying $u$ by $f\rho$ and integrating over $\Omega$ gives $Gb=0$, since $\langle f_j,h\rangle_\rho=0$, $j=1,\dots,n_\theta$. As $G\succ0$, $b=0$, which contradicts $\|b\|_2\geq\min\{1,\alpha\}$. \qed
\end{pf}

Theorem~\ref{thm:sm.conv} holds for every kernel satisfying Assumption~\ref{ass:ok.kernel} and every input sequence that becomes dense in $\Omega$ with vanishing noise bound. It does not extend to $\mathcal{S}(\mathcal{D})$ in general, as shown next.

\begin{proposition}[unconstrained feasible set]\label{prop:sm.kme}
Let Assumptions~\ref{ass:ok.geometry}, \ref{ass:ok.kernel}, and~\ref{ass:sm.info} hold, let $f_j\in\mathcal{H}$, $j=1,\dots,n_\theta$, and let $\|\Delta_\perp\|_\mathcal{H}<\beta$. Let $G_\mathcal{H}\doteq[\langle f_i,f_j\rangle_\mathcal{H}]_{i,j=1}^{n_\theta}$, $a_\mathcal{H}\doteq G_\mathcal{H}^{-1}[\langle f_1,\Delta_\perp\rangle_\mathcal{H},\dots,\langle f_{n_\theta},\Delta_\perp\rangle_\mathcal{H}]^\top$, and $s_\mathcal{H}\doteq(\beta^2-\|\Delta_\perp-f^\top a_\mathcal{H}\|_\mathcal{H}^2)^{1/2}$. Then, $s_\mathcal{H}>0$ and, for every data set satisfying Assumption~\ref{ass:sm.info},
\begin{equation}
\mathcal{S}(\mathcal{D})\supseteq\bigl\{\theta_\rho+a:\ (a-a_\mathcal{H})^\top G_\mathcal{H}(a-a_\mathcal{H})\leq s_\mathcal{H}^2\bigr\}.
\label{eqn:sm.kme}
\end{equation}
In particular,
\[
\sup_{\theta,\theta'\in\mathcal{S}(\mathcal{D})}\|\theta-\theta'\|_2\geq2\,s_\mathcal{H}\,\lambda_{\min}(G_\mathcal{H})^{-1/2}.
\]
\end{proposition}

\begin{pf}
If $c^\top G_\mathcal{H}c=\|f^\top c\|_\mathcal{H}^2=0$, then $f^\top c=0$ on $\mathcal{X}$, hence $c^\top Gc=0$ and $c=0$, since $G\succ0$; thus, $G_\mathcal{H}\succ0$. Since $\Delta_\perp\in\mathcal{H}$ by Assumption~\ref{ass:sm.info}(ii), $a_\mathcal{H}$ is the solution of the normal equations of the minimization of $\|\Delta_\perp-f^\top a\|_\mathcal{H}$ over $a\in\mathbb{R}^{n_\theta}$, so that $\|\Delta_\perp-f^\top a_\mathcal{H}\|_\mathcal{H}\leq\|\Delta_\perp\|_\mathcal{H}<\beta$, i.e., $s_\mathcal{H}>0$. Moreover, $\Delta_\perp-f^\top a_\mathcal{H}$ is $\mathcal{H}$-orthogonal to each $f_j$; therefore, for every $a\in\mathbb{R}^{n_\theta}$,
\[
\begin{aligned}
\|\Delta_\perp-f^\top a\|_\mathcal{H}^2={}&\|\Delta_\perp-f^\top a_\mathcal{H}\|_\mathcal{H}^2\\
&+(a-a_\mathcal{H})^\top G_\mathcal{H}(a-a_\mathcal{H}).
\end{aligned}
\]
Let $a$ belong to the ellipsoid in~\eqref{eqn:sm.kme}, and let $\delta\doteq\Delta_\perp-f^\top a\in\mathcal{H}$, so that $\|\delta\|_\mathcal{H}^2\leq\|\Delta_\perp-f^\top a_\mathcal{H}\|_\mathcal{H}^2+s_\mathcal{H}^2=\beta^2$. By Assumption~\ref{ass:sm.info}(i), $Y-F(x)(\theta_\rho+a)=[\Delta_\perp-f^\top a]_\mathcal{D}+e=[\delta]_\mathcal{D}+e$ for some $e\in\mathcal{E}$, i.e., $Y-F(x)(\theta_\rho+a)\in\mathcal{W}$, and $\theta_\rho+a\in\mathcal{S}(\mathcal{D})$. The last bound follows by taking $a=a_\mathcal{H}\pm s_\mathcal{H}\lambda_{\min}(G_\mathcal{H})^{-1/2}v$, where $v$ is a unit eigenvector of $G_\mathcal{H}$ associated with $\lambda_{\min}(G_\mathcal{H})$: both points lie on the boundary of the ellipsoid, and their distance is $2s_\mathcal{H}\lambda_{\min}(G_\mathcal{H})^{-1/2}$. \qed
\end{pf}

The lower bound in Proposition~\ref{prop:sm.kme} does not depend on $T$, $\mathcal{E}$, or the inputs, and applies, e.g., to polynomial $f$ and Mat\'ern kernels on a bounded Lipschitz domain $\mathcal{X}$, whose RKHS is norm-equivalent to a Sobolev space~\cite{wendland2004scattered}.
Since the proof perturbs only the entries associated with the $f_j$ in $\mathcal{H}$, if only some of them belong to $\mathcal{H}$, \eqref{eqn:sm.kme} holds with $a$, $G_\mathcal{H}$, and $a_\mathcal{H}$ restricted to these entries.
The ellipsoid is centered at the parameter of the member of \eqref{eqn:ok.ambiguity} whose correction is $\mathcal{H}$-orthogonal to $\mathcal{F}$, i.e., the condition satisfied by the KME solution in Remark~\ref{rem:ok.implicit}. 
Hence, if $\mathcal{F}\subset\mathcal{H}$ and $\|\Delta_\perp\|_\mathcal{H}<\beta$, $\mathcal{S}(\mathcal{D})$ does not shrink as the inputs become dense and the noise vanishes, whereas, under the same information, $\mathcal{S}_\perp(\mathcal{D})$ shrinks to $\theta_\rho$ by Theorem~\ref{thm:sm.conv}.

\section{Implementation}\label{sec:impl}
The analysis of Section~\ref{sec:orthker} relies on functional-analytic arguments, but its practical use is simple. By Theorem~\ref{thm:ok.embedding}, the estimate is obtained as in Section~\ref{sec:kerneltheory}, with $\mathbf{K}$ replaced by $\mathbf{K}_\perp$, and the only additional ingredients are the representers $\eta$ and the matrix $H$, which depend on the basis $f$, the reference geometry $(\Omega,\rho)$, and the kernel $\kappa$, but not on the data. Section~\ref{sec:impl.design} discusses the design choices, Section~\ref{sec:impl.quad} describes how $\eta$ and $H$ are computed, and Section~\ref{sec:impl.alg} summarizes the step-by-step procedure.

\subsection{Design choices}\label{sec:impl.design}

\subsubsection{Operating domain and weight} The set $\Omega$ is the region where the physical parameters are to be interpreted, e.g., the operating envelope of the system. By Corollary~\ref{cor:ok.bound}, the error on $\theta_\rho$ is controlled by the reconstruction error of $\bar g$ on $\Omega$. Hence, $\Omega$ should be covered by the identification inputs. Inputs outside $\Omega$ are also admissible, since $\kappa_\perp$ is defined on $\mathcal{X}$. The uniform weight $\rho\equiv1$ is a natural default, whereas a nonuniform $\rho$ emphasizes the regions where the physical model is most relevant. Since $\theta_\rho$ depends on $(\Omega,\rho)$, these choices are fixed before identification and should be reported together with the estimates.

\subsubsection{Physical basis} Assumption~\ref{ass:ok.geometry}(iv) can be checked on $G$, computed, like $\eta$ and $H$, in closed form or by quadrature (Section~\ref{sec:impl.quad}). A small $\lambda_{\min}(G)$ indicates nearly redundant parameters and, by \eqref{eqn:ok.bound}, amplifies the error on $\theta_\rho$; in this case, $f$ should be reparametrized.

\subsubsection{Kernel and regularization} By Theorem~\ref{thm:ok.kperp}, any kernel satisfying Assumption~\ref{ass:ok.kernel} can be used, and $H$ need not be invertible, since \eqref{eqn:ok.kperp} involves its pseudoinverse. As in standard kernel ridge regression, $\kappa$ encodes prior knowledge on the regularity of the discrepancy, e.g., through Gaussian or Mat\'ern kernels. The hyperparameters of $\kappa$ and the weight $\gamma$ can be selected by standard methods, such as cross-validation or marginal-likelihood maximization, with $\mathbf{K}_\perp$ in place of $\mathbf{K}$. In contrast with Problem~\eqref{eqn:optprob.theta} (see Remark~\ref{rem:ok.implicit}), these choices affect the estimate $\theta^\star_\perp$ but not its target $\theta_\rho$, which depends only on $(\bar g,f,\Omega,\rho)$. Since $\eta$ and $H$ depend on $\kappa$, they must be recomputed for each candidate value of the kernel hyperparameters, but not of $\gamma$. If $\mathcal{H}$ is finite-dimensional, e.g., for polynomial kernels, then $\dim\mathcal{H}_\perp=\dim\mathcal{H}-\operatorname{rank}H$, since $\operatorname{span}\{\eta_1,\dots,\eta_{n_\theta}\}$ has dimension $\operatorname{rank}H$; hence, $\dim\mathcal{H}$ must exceed $n_\theta$ sufficiently, or no admissible correction is left. The bound $\beta$ and the set $\mathcal{E}$ of Section~\ref{sec:sm} are prior information; their selection from data is not addressed here.

\subsection{Moment representers and Gram matrix}\label{sec:impl.quad}

By \eqref{eqn:ok.eta}--\eqref{eqn:ok.H},
\[
H=\int_\Omega f(x)\,\eta(x)^\top\rho(x)\,dx;
\]
hence, once $\eta$ is available, $H$ is a single integral of a known function, and the computation reduces to the evaluation of the integral in \eqref{eqn:ok.eta}.

\subsubsection{Closed forms} Given an interval $\Omega$, $\rho\equiv1$, and a polynomial basis $f$, both $\eta$ and $H$ admit closed forms for the most commonly used kernels: they involve only polynomials for polynomial kernels, polynomials and exponentials for Mat\'ern kernels of half-integer order, and polynomials, exponentials, and the error function for Gaussian kernels. Moreover, for box-shaped $\Omega$, product kernels, product weights, and monomial bases, the integrals factorize into one-dimensional ones. These derivations are reported in Section~\ref{app:closedform} of the Supplementary material.

\subsubsection{Quadrature} As an alternative to closed forms, the integrals can be evaluated by a quadrature rule with nodes $\xi_1,\dots,\xi_M\in\Omega$ and weights $w_1,\dots,w_M>0$, i.e.,
\[
\int_\Omega g(x)\rho(x)\,dx\approx\sum_{m=1}^Mw_m\,g(\xi_m),
\]
where the weights account for $\rho$. Let
\begin{gather*}
\Phi\doteq[f(\xi_1),\dots,f(\xi_M)]^\top,\, N\doteq\operatorname{diag}(w_1,\dots,w_M),\\
k_\Xi(x)\doteq[\kappa(\xi_1,x),\dots,\kappa(\xi_M,x)]^\top,\, (\mathbf{K}_\Xi)_{ml}\doteq\kappa(\xi_m,\xi_l).
\end{gather*}
Applying the rule to the integral in \eqref{eqn:ok.eta} and to both integrals in \eqref{eqn:ok.H} gives the approximations
\begin{equation}
\eta_M(x)\doteq\Phi^\top N\,k_\Xi(x),
\qquad
H_M\doteq\Phi^\top N\mathbf{K}_\Xi N\Phi
\label{eqn:impl.quad}
\end{equation}
of $\eta(x)$ and $H$, respectively. Replacing $\eta$ and $H$ with $\eta_M$ and $H_M$ in \eqref{eqn:ok.kperp} yields the orthogonalized kernel associated with the discrete inner product $\sum_{m=1}^Mw_mg(\xi_m)q(\xi_m)$. Indeed, provided that $\Phi^\top N\Phi\succ0$, Lemma~\ref{lem:ok.riesz} and Theorem~\ref{thm:ok.kperp} hold, with the same proofs, when the integrals with respect to $\rho(x)\,dx$ are replaced by the quadrature sum. Hence, the resulting corrections are exactly orthogonal to $\mathcal{F}$ in this discrete sense, and approximately orthogonal in the sense of \eqref{eqn:ok.inner}. The choice of the rule and its convergence are discussed in Section~\ref{app:closedform} of the Supplementary material. Finally, if $H_M$ is ill-conditioned, a truncated pseudoinverse can be used, which relaxes the constraints along the discarded directions. In the following, $\eta$ and $H$ denote either their closed forms or their approximations \eqref{eqn:impl.quad}.

\subsubsection{Verification} Regardless of how $\eta$ and $H$ are obtained, the computation can be checked through the physical moments of $\kappa_\perp$. By Theorem~\ref{thm:ok.kperp}(iv), we have $\int_\Omega f(\xi)\kappa_\perp(\xi,x)\rho(\xi)\,d\xi=0$ for every $x\in\mathcal{X}$; evaluated at a set of test points with a rule finer than the one used for $\eta$ and $H$, this integral measures how accurately the constraints in \eqref{eqn:ok.Hperp} are enforced. When closed forms are used, it is at roundoff level.

\subsection{Procedure}\label{sec:impl.alg}

Algorithm~\ref{alg:impl} summarizes the procedure for given $\kappa$ and $\gamma$. Steps~1--2 depend only on $(f,\Omega,\rho,\kappa)$ and are performed offline, once for each value of the kernel hyperparameters (Section~\ref{sec:impl.design}); with quadrature, they require $O(M^2)$ kernel evaluations. Steps~3--5 compute the estimate and have the cost of standard kernel ridge regression, dominated by the $O(T^3)$ inversion in \eqref{eqn:ok.Kperp}, plus $O(T)$ evaluations of $\eta$ for $E(x)$. Steps~6--7 follow identification and serve prediction and diagnosis. The feasible sets of Section~\ref{sec:sm} are not needed by the procedure; given $\beta$ and $\mathcal{E}$, they are computed by the second-order cone programs of Section~\ref{sec:sm.fps}, with $\kappa_\perp$ fixed at the selected hyperparameters.

\begin{algorithm}[t]
\caption{Identification with the orthogonalized kernel}\label{alg:impl}
\begin{algorithmic}[1]
\Require data $\mathcal{D}$; physical basis $f$; operating domain $\Omega$ and weight $\rho$; kernel $\kappa$ and $\gamma>0$.
\State \textit{Offline.} Obtain $\eta(\cdot)$ and $H$, in closed form or by \eqref{eqn:impl.quad}, and check that $G\succ0$.
\State \textit{Offline.} Compute the (possibly truncated) pseudoinverse $H^\dagger$.
\State Form $F(x)$, $Y$, and $\mathbf{K}$; compute $E(x)$, with rows $\eta(x_t)^\top$, $t=1,\dots,T$.
\State Compute $\mathbf{K}_\perp$ and $\Psi_\perp$ by \eqref{eqn:ok.Kperp}.
\State Compute $\theta^\star_\perp$ by \eqref{eqn:ok.theta}, $\omega^\star_\perp=\Psi_\perp\bigl(Y-F(x)\theta^\star_\perp\bigr)$, and $\zeta\doteq H^\dagger E(x)^\top\omega^\star_\perp\in\mathbb{R}^{n_\theta}$.
\State \textit{Prediction.} For $x\in\mathcal{X}$, $\delta^\star_\perp(x)=\sum_{j=1}^T\omega^\star_{\perp,j}\kappa(x,x_j)-\eta(x)^\top\zeta$, and the predicted output is $f(x)^\top\theta^\star_\perp+\delta^\star_\perp(x)$.
\State \textit{Diagnostic (optional).} Compute $r\doteq\|\delta^\star_\perp\|_\rho/\|\hat g_\perp\|_\rho$.
\Ensure $\theta^\star_\perp$, $\delta^\star_\perp$, and, optionally, $r$.
\end{algorithmic}
\end{algorithm}

Step~6 follows from \eqref{eqn:ok.delta} and \eqref{eqn:ok.kperp}, since
\[
\textstyle\sum_{j=1}^T\omega^\star_{\perp,j}\eta(x_j)=E(x)^\top\omega^\star_\perp;
\]
hence, the vector $\zeta$ is computed once, and each prediction only requires the evaluation of $\kappa(x,x_j)$, $j=1,\dots,T$, and of $\eta(x)$.

Step~7 quantifies the role of the correction. Since $\delta^\star_\perp\perp_\rho\mathcal{F}$ by Theorem~\ref{thm:ok.embedding}, $\langle f^\top\theta^\star_\perp,\delta^\star_\perp\rangle_\rho=0$, and hence
\[
\|\hat g_\perp\|_\rho^2=\|f^\top\theta^\star_\perp\|_\rho^2+\|\delta^\star_\perp\|_\rho^2 .
\]
Therefore, for $\hat g_\perp\neq0$, $r\in[0,1]$, and $r^2$ is the fraction of $\|\hat g_\perp\|_\rho^2$ that lies outside $\mathcal{F}$. If $\eta$ and $H$ are computed by \eqref{eqn:impl.quad} and the norms are evaluated with the same rule, this identity holds exactly, since the correction is then exactly orthogonal in the corresponding discrete sense. A value of $r$ close to one indicates that, on $\Omega$, the fitted map is explained mostly by the correction rather than by the physical model.

\section{Numerical results}\label{sec:num}
This section illustrates the results of Sections~\ref{sec:orthker} and~\ref{sec:sm} on a static example and on a simulated mass-spring-damper. Three estimates of the physical parameters are compared: the least-squares (LS) estimate
\[
\bigl(F(x)^\top F(x)\bigr)^{-1}F(x)^\top Y,
\]
obtained from the physical model alone, the KME estimate~\eqref{eqn:thmlincf}, and the estimate~\eqref{eqn:ok.theta}, referred to as orthogonal KME (OKME).

For scalar inputs, the kernel is the Mat\'ern kernel of order $3/2$,
\[
\kappa(x,x')=\Bigl(1+\frac{\sqrt{3}\,|x-x'|}{\ell}\Bigr)\exp\Bigl(-\frac{\sqrt{3}\,|x-x'|}{\ell}\Bigr),
\]
with length-scale $\ell>0$; for vector inputs, it is the product of such factors, each with its own length-scale. Its amplitude is absorbed in $\gamma$. Unless stated otherwise, the length-scales and $\gamma$ maximize the restricted marginal likelihood (REML)~\cite{patterson1971recovery} of the model in which $\delta$ is a Gaussian process with covariance proportional to $\kappa$ and $\theta$ has a noninformative prior, with $\mathbf{K}$ for KME and $\mathbf{K}_\perp$ for OKME (Section~\ref{sec:impl.design}). The vector $\eta$ and the matrix $H$ are computed in closed form when the factors of $f$ are monomials, and by Gauss--Legendre quadrature otherwise; in all cases, the verification of Section~\ref{sec:impl.quad} returns relative residuals below $10^{-10}$.

The noise is uniformly distributed in $[-\varepsilon,\varepsilon]$. The feasible sets are computed for the box $\mathcal{E}=\{e:\|e\|_\infty\leq\varepsilon\}$ and $\beta=1.5\,\|\Delta_\perp\|_\mathcal{H}$, with the kernel fixed at the selected hyperparameters and $\|\Delta_\perp\|_\mathcal{H}$ approximated by minimum-norm interpolation on a fine grid; the second-order cone programs of Section~\ref{sec:sm.fps} are solved with MOSEK.

\subsection{Static example}\label{sec:num.static}
Consider the map~\eqref{eqn:system.comp.lin} with $x\in\mathcal{X}=\Omega=[-1,1]$, $\rho\equiv1$, $f(x)=[1,\ x]^\top$, $\bar\theta=[0.5,\ 1]^\top$, $\Delta(x)=0.6\tanh(4x)+0.3\cos(3x)$, and $\varepsilon=0.05$, which corresponds to a signal-to-noise ratio of about $30$~dB. The canonical parameter is $\theta_\rho=[0.514,\ 1.854]^\top$, since the odd part of $\Delta$ mainly increases the slope. The RKHS of the Mat\'ern kernel of order $3/2$ on $\Omega$ is norm-equivalent to the Sobolev space of order~$2$~\cite{wendland2004scattered}; hence, $\mathcal{F}\subset\mathcal{H}$ and $\Delta_\perp\in\mathcal{H}$.

\begin{figure}[t]
\centering
\includegraphics{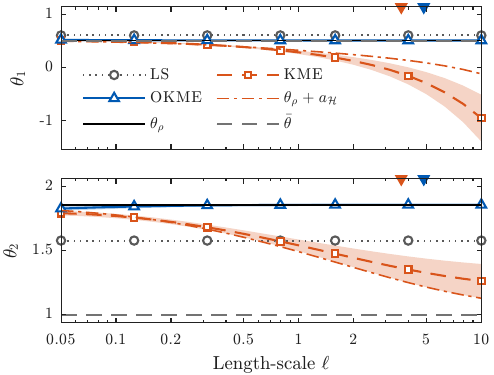}
\caption{Estimates of $\theta_1$ (top) and $\theta_2$ (bottom) versus $\ell$: mean over $100$ data sets and $\pm$ one standard deviation (shaded). Triangles: median length-scales selected by REML. Top: $\bar\theta_1$ and the first entry of $\theta_\rho$ nearly coincide.}
\label{fig:num.static.ell}
\end{figure}

For $100$ data sets of $T=200$ samples, with inputs drawn from the density $(1+x)/2$, which differs from $\rho$, the LS, KME, and OKME estimates are computed for $24$ values of $\ell\in[0.05,10]$, with $\gamma$ selected by REML for each $\ell$; Fig.~\ref{fig:num.static.ell} shows their means as functions of $\ell$. The mean KME estimate moves from $[0.51,\ 1.79]^\top$ at $\ell=0.05$ to $[-0.95,\ 1.26]^\top$ at $\ell=10$, and is close to $[-0.15,\ 1.35]^\top$ at $\ell\approx4$, near the median length-scale selected by REML for KME, $\ell=3.6$. This is not a finite-sample effect: by Remark~\ref{rem:ok.implicit}, if the fitted map coincided with $\bar g$, its parameter would be $\theta_\rho+a_\mathcal{H}$, with $a_\mathcal{H}$ as in Proposition~\ref{prop:sm.kme}; since $a_\mathcal{H}$ depends on $\ell$ through the inner product of $\mathcal{H}$, $\theta_\rho+a_\mathcal{H}$ follows the same trend (dash-dotted line in Fig.~\ref{fig:num.static.ell}). The mean OKME estimate differs from $\theta_\rho$ by less than $0.03$ for every $\ell$, and by less than $0.005$ for $\ell\geq0.3$, with standard deviations below $0.02$: the kernel affects the estimate, but not its target. The LS estimate, $[0.61,\ 1.58]^\top$, does not depend on $\ell$ and is biased, since it weights the approximation error by the input density instead of $\rho$.

\begin{figure}[t]
\centering
\includegraphics{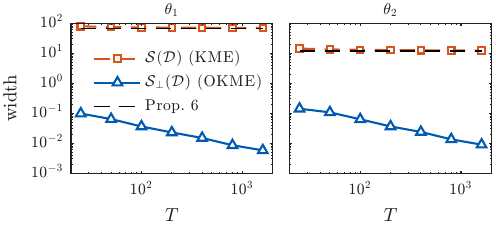}
\caption{Widths of $\mathcal{S}(\mathcal{D})$ and $\mathcal{S}_\perp(\mathcal{D})$ along $\theta_1$ (left) and $\theta_2$ (right) versus $T$, with inputs uniform on $\Omega$ (median over $10$ data sets). Dashed: lower bound from Proposition~\ref{prop:sm.kme}.}
\label{fig:num.static.fps}
\end{figure}

\begin{figure}[t]
\centering
\includegraphics{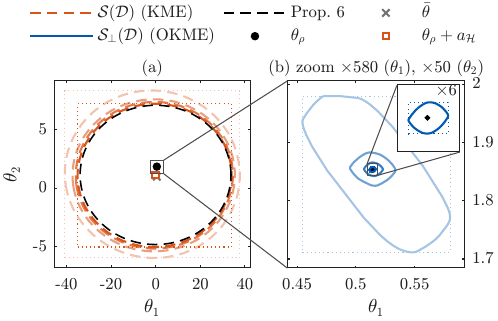}
\caption{Feasible sets for the first $T$ samples of one input sequence, $T=25$, $100$, $400$, $1600$ (light to dark): (a) $\mathcal{S}(\mathcal{D})$ and the ellipsoid~\eqref{eqn:sm.kme}; (b) $\mathcal{S}_\perp(\mathcal{D})$, with a $\times6$ inset around $\theta_\rho$. Sets are drawn as outer polygons from $64$ support directions (area excess below $2\%$); dotted: extents for $T=25$ and $1600$. The square in (a) marks the position of (b), not its size.}
\label{fig:num.static.plane}
\end{figure}

The feasible sets are computed with the kernel fixed at $\ell=8.60$, selected by REML with $\mathbf{K}_\perp$ on an additional data set, so that $\|\Delta_\perp\|_\mathcal{H}=31.5$, in two settings: \emph{full coverage}, with inputs uniform on $\Omega$, and \emph{partial coverage}, with inputs uniform on $[-1,\,0.2]$. For each setting and each $T\in\{25,50,100,200,400,800,1600\}$, $10$ data sets are generated, and $\theta_\rho\in\mathcal{S}_\perp(\mathcal{D})$ in all $140$ cases. With full coverage (Fig.~\ref{fig:num.static.fps}), the widths of $\mathcal{S}_\perp(\mathcal{D})$ along $\theta_1$ and $\theta_2$ decrease from $0.097$ and $0.141$ to $0.0058$ and $0.0089$, consistently with Theorem~\ref{thm:sm.conv}, although $\varepsilon$ is fixed here: the uniform noise takes values arbitrarily close to $\pm\varepsilon$, so that the noise constraint becomes tight as the data become dense. The widths of $\mathcal{S}(\mathcal{D})$ only decrease from $78.8$ and $14.2$ to $68.9$ and $12.2$, above the lower bounds $67.6$ and $11.9$ given by the projections of the ellipsoid~\eqref{eqn:sm.kme}. Fig.~\ref{fig:num.static.plane} shows the two sets in the parameter plane for the first $T$ samples of one input sequence with full coverage, so that the sets are nested: $\mathcal{S}(\mathcal{D})$ barely changes with $T$ and contains the ellipsoid~\eqref{eqn:sm.kme}, centered at $\theta_\rho+a_\mathcal{H}$, whereas $\mathcal{S}_\perp(\mathcal{D})$ shrinks around $\theta_\rho$. In contrast, with partial coverage, the fill distance in $\Omega$ does not vanish, and the widths of $\mathcal{S}_\perp(\mathcal{D})$ remain about $1$ and $2.2$, since the data carry no information on $\bar g$ over $(0.2,\,1]$.

\subsection{Mass-spring-damper}\label{sec:num.msd}
In the second example, a unit mass with linear spring, viscous damper, and input gain is subject to a cubic stiffness and a smooth Coulomb friction, both neglected by the physical model. The data are generated by the discrete-time model
\[
\begin{aligned}
v_{t+1}&=v_t+T_s\bigl(-kp_t-cv_t+bu_t+\Delta(p_t,v_t)+e_t\bigr),\\
p_{t+1}&=p_t+T_sv_{t+1},
\end{aligned}
\]
with $T_s=0.05$, $k=4$, $c=0.4$, $b=1$, and $\Delta(p,v)=-k_3p^3-F_c\tanh(v/v_s)$, $k_3=2$, $F_c=0.3$, $v_s=0.2$. With $y_t\doteq(v_{t+1}-v_t)/T_s$ and $x_t\doteq(p_t,v_t,u_t)$, this is the map~\eqref{eqn:system.comp.lin} with $f(x)=[p,\ v,\ u]^\top$, and $\bar\theta=[-k,\ -c,\ b]^\top$. The operating domain is $\Omega=[-1,1]\times[-2,2]\times[-3,3]$, with $\rho\equiv1$. The canonical parameter $\theta_\rho$ corresponds to $k=5.2$, $c=0.623$, and $b=1$, i.e., to the linear spring and damper that best approximate in $L^2_\rho$ the restoring and dissipative forces of the system on $\Omega$; the input gain is unchanged, since $\Delta$ does not depend on $u$. The kernel is the product Mat\'ern kernel on $(p,v)$. As $\kappa$ does not depend on $u$ and $\int_{-3}^{3}u\,du=0$, the third entry of $\eta$ vanishes and $H$ is singular; as discussed after Theorem~\ref{thm:ok.kperp}, the pseudoinverse in~\eqref{eqn:ok.kperp} accounts for this exactly.

Five identification experiments with $T=1000$ samples are considered, which differ in the input: random-phase multisines with flat spectrum on $[0.1,4]$, $[0.05,1]$, $[1.4,2.6]$, and $[2.5,6]$~rad/s (E1, E2, E3, and E5, respectively; the natural frequency is $2$~rad/s), and a random binary signal with a clock period of six samples (E4). The amplitude of each input is the smallest value, in a geometric sequence, for which the noise-free trajectory visits at least $95\%$ of the cells of an $8\times8$ grid on the projection of $\Omega$ onto the $(p,v)$ plane. Hence, all experiments (except for E2) explore the operating domain, with different distributions on it. The slow input of E2 attains at most $83\%$, with large displacements, and only $19\%$ of its samples lie in $\Omega$. The noise bound $\varepsilon$ is set so that the signal-to-noise ratio, defined as the ratio between the sample variance of the noise-free $y_t$ and the noise variance $\varepsilon^2/3$, is $40$ or $20$~dB. For each experiment and noise level, $100$ data sets are generated with new realizations of the input and of the noise. Prediction is assessed on a noise-free validation record of $1000$ samples, with a multisine input on $[0.1,4]$~rad/s of a different realization, through the normalized root-mean-square error (NRMSE) of the one-step prediction of $y_t$ and of the position simulated by the identified model, including the estimated correction for KME and OKME.

\begin{figure}[t]
\centering
\includegraphics{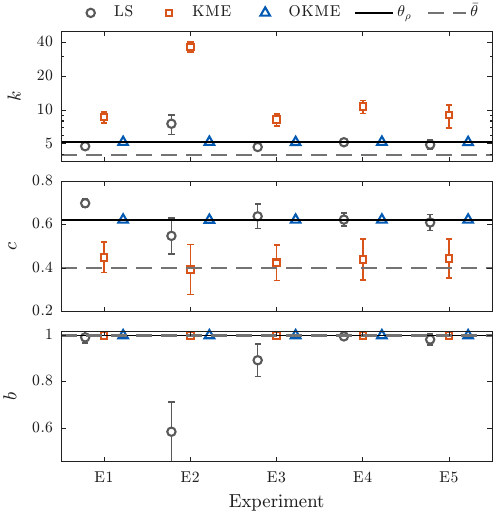}
\caption{Estimates of $k$ (top, logarithmic scale), $c$ (middle), and $b$ (bottom) in the five experiments at $40$~dB: mean and $\pm$ one standard deviation over $100$ data sets.}
\label{fig:num.msd.exp}
\end{figure}

Fig.~\ref{fig:num.msd.exp} shows the mean and the standard deviation of the estimates in each experiment at $40$~dB. The OKME estimates of $k$ and $c$ lie between $5.17$ and $5.20$ and at $0.623$, respectively, in all experiments, and between $5.10$ and $5.20$ and between $0.624$ and $0.628$ at $20$~dB. The KME estimates of $k$ range from $8.2$ to $36.6$, with the largest value in E2, and those of $c$ from $0.39$ to $0.45$; hence, they depend on the experiment and differ from $\theta_\rho$, as expected from Remark~\ref{rem:ok.implicit}, since the decomposition selected by Problem~\eqref{eqn:optprob.theta} depends on the data. The LS estimates also vary with the experiment, since they depend on the distribution of the data, and underestimate the input gain in E2 and E3 ($0.58$ and $0.89$). The input gain is estimated within $0.1\%$ by both KME and OKME in all experiments.

Over all experiments and data sets, the root-mean-square error of OKME with respect to $\theta_\rho$ is $0.7\%$ and $0.5\%$ of the corresponding entries on $k$ and $c$ at $40$~dB, and $1.9\%$ and $3.3\%$ at $20$~dB. Table~\ref{tab:num.msd} reports the median NRMSE on the validation record, i.e., the root-mean-square prediction error divided by the standard deviation of the predicted signal. OKME and KME predict comparably, with NRMSE an order of magnitude below that of LS at $40$~dB and four to five times below it at $20$~dB, and that of OKME exceeding that of KME by at most $8\%$; hence, constraining the correction to $\mathcal{H}_\perp$ fixes the physical parameter without a significant loss in the description of the data.

\begin{table}[t]
\caption{Mass-spring-damper: NRMSE on the validation record (one-step prediction and simulation) and median widths along $k$ and $c$ of $\mathcal{S}(\mathcal{D})$ (KME row) and $\mathcal{S}_\perp(\mathcal{D})$ (OKME row).}
\label{tab:num.msd}
\centering
\begin{tabular}{llrrrr}
\toprule
 & & \multicolumn{2}{c}{NRMSE (\%)} & \multicolumn{2}{c}{Width} \\
\cmidrule(lr){3-4}\cmidrule(lr){5-6}
SNR & Method & One-step & Sim. & $k$ & $c$ \\
\midrule
\multirow{3}{*}{40 dB} & LS   & 21.95 & 44.97 & --   & -- \\
                       & KME  & 1.58  & 2.04  & 54.2 & 26.0 \\
                       & OKME & 1.66  & 2.21  & 0.11 & 0.04 \\
\midrule
\multirow{3}{*}{20 dB} & LS   & 21.93 & 44.82 & --   & -- \\
                       & KME  & 5.63  & 8.16  & 51.2 & 23.9 \\
                       & OKME & 5.66  & 8.52  & 0.45 & 0.20 \\
\bottomrule
\end{tabular}
\end{table}

The feasible sets are computed for the first $20$ data sets of each experiment and noise level, with the kernel selected by REML with $\mathbf{K}_\perp$ and $\|\Delta_\perp\|_\mathcal{H}$ evaluated for the restriction of $\mathcal{H}$ to the box spanned by the data in $(p,v)$. In agreement with Proposition~\ref{prop:sm.fps}(i), $\theta_\rho\in\mathcal{S}_\perp(\mathcal{D})$ in all $200$ cases. The median widths of $\mathcal{S}_\perp(\mathcal{D})$ along $k$ and $c$, reported in Table~\ref{tab:num.msd}, are more than two orders of magnitude smaller than those of $\mathcal{S}(\mathcal{D})$, which remain above $45$ and $22$ in all experiments at $40$~dB, and above $37$ and $16$ at $20$~dB, in agreement with Theorem~\ref{thm:sm.conv} and Proposition~\ref{prop:sm.kme}. The widths of $\mathcal{S}_\perp(\mathcal{D})$ are largest for E2, three to four times the median over the other experiments, consistently with the small fraction of its samples in $\Omega$. For E1, with $\beta$ equal to $1$ and $2$ times $\|\Delta_\perp\|_\mathcal{H}$, the widths of $\mathcal{S}_\perp(\mathcal{D})$ along $k$ are $0.045$ and $0.11$ at $40$~dB, and those of $\mathcal{S}(\mathcal{D})$ are $19.5$ and $72.6$.

The re-labeled estimate of Remark~\ref{rem:ok.geometries} is not reported in Table~\ref{tab:num.msd}: it has the same fitted map as KME, and hence the same NRMSE, whereas its target $\theta_\rho$ is fixed by the reference geometry and its feasible set is $\mathcal{S}_\perp(\mathcal{D})$, defined by the orthogonalized kernel. Its relative RMSE is $0.66\%$ and $0.49\%$ on $k$ and $c$ at $40$~dB, statistically indistinguishable from OKME in this setting. In this example, $f_1=p$ and $f_2=v$ belong to $\mathcal{H}$, and $f_3=u$ is $\rho$-orthogonal to it; hence, $\Pi_\rho\delta^\star\in\operatorname{span}\{f_1,f_2\}$, and the correction $\delta^\star-\Pi_\rho\delta^\star$ of the re-labeled estimate belongs to $\mathcal{H}$, as that of OKME. To examine a physical basis that is not contained in $\mathcal{H}$, for which this fails in general, Coulomb friction is included both in the system and in the physical model: the term $-F_c\operatorname{sign}(v_t)$, with $F_c=0.3$, is added to the acceleration, $f(x)=[p,\ v,\ u,\ \operatorname{sign}(v)]^\top$, $\bar\theta=[-k,\ -c,\ b,\ -F_c]^\top$, and $\Delta(p,v)=-k_3p^3-d\,p^2v$, $d=1$, which represents a position-dependent damping. Then, $\theta_\rho$ corresponds to $k=5.2$, $c=0.733$, $b=1$, and $F_c=0.3$, and $\Delta_\perp(p,v)=-k_3(p^3-3p/5)-d(p^2-1/3)v$ is continuous and belongs to $\mathcal{H}$, whereas $\operatorname{sign}(v)$ does not. For the five experiments, with amplitudes recalibrated as above, both noise levels, and $20$ data sets each, the length-scales are selected by REML with $\mathbf{K}_\perp$ and kept for all estimators, and the noise set is the Euclidean ball of radius $\mu=\sqrt{T}\varepsilon$, which contains the box. Table~\ref{tab:num.coulomb} reports the worst-case errors of the linear estimators, given by Proposition~\ref{prop:sm.acc}(ii), divided by the optimal worst-case errors $\operatorname{err}^\star_j$, computed by the second-order cone programs of Proposition~\ref{prop:sm.acc}.

\begin{table}[t]
\caption{Mass-spring-damper with Coulomb friction: worst-case errors divided by $\operatorname{err}^\star_j$ (median over experiments and data sets), with $\gamma$ minimizing the worst-case error for each $j$ and with $\gamma$ selected by REML.}
\label{tab:num.coulomb}
\centering
\setlength{\tabcolsep}{2.5pt}
\begin{tabular}{llrrrrrr}
\toprule
 & & \multicolumn{3}{c}{40 dB} & \multicolumn{3}{c}{20 dB} \\
\cmidrule(lr){3-5}\cmidrule(lr){6-8}
$\gamma$ & Method & $k$ & $c$ & $F_c$ & $k$ & $c$ & $F_c$ \\
\midrule
\multirow{4}{*}{Best} & OKME           & 1.00 & 1.00 & 1.00 & 1.00 & 1.00 & 1.00 \\
                      & Re-labeled KME & 1.01 & 1.07 & 1.08 & 1.00 & 1.04 & 1.04 \\
                      & KME            & 5.37 & 4.23 & 1.69 & 1.49 & 1.26 & 1.09 \\
                      & LS             & 19.9 & 25.2 & 39.8 & 3.01 & 2.44 & 3.45 \\
\midrule
\multirow{3}{*}{REML} & OKME           & 1.03 & 1.02 & 1.01 & 1.02 & 1.02 & 1.01 \\
                      & Re-labeled KME & 1.05 & 1.12 & 1.10 & 1.02 & 1.06 & 1.06 \\
                      & KME            & 95.0 & 34.0 & 1.87 & 14.3 & 4.98 & 1.44 \\
\bottomrule
\end{tabular}
\end{table}

With the best $\gamma$, OKME attains $\operatorname{err}^\star_j$ within $0.1\%$ in all cases, in agreement with Proposition~\ref{prop:sm.acc}(iv), whereas the re-labeled estimate does not. Its median excess is about $1\%$ on $k$ and larger on $c$ and $F_c$, the parameters of $v$ and of $\operatorname{sign}(v)$, which are correlated on $\Omega$; it depends strongly on the experiment: on $c$ and $F_c$ at $40$~dB, it is about $2\%$ for E5, $20\%$ for E1, and a factor of $2.8$ and $3.8$, respectively, for E2. Hence, when the physical basis is not contained in $\mathcal{H}$, re-labeling the unconstrained fit does not attain the best achievable accuracy, while restricting the correction class does.

\section{Conclusion}\label{sec:concl}
This paper extended kernel-based model embedding with a reference geometry, defined by an operating domain and a weight, which fixes the part of the input--output map that is regarded as physical. The resulting canonical parameter depends on neither the kernel nor the data, and the orthogonalized kernel restricts the correction to the functions that are orthogonal to the physical model in this geometry. The orthogonal estimator retains the closed form and the computational structure of KME for models affine in the parameters, with the kernel matrix as the only modification, and adds two properties: a target that is independent of the kernel, and a feasible parameter set that converges to the canonical parameter as the inputs become dense in the operating domain and the noise vanishes. Moreover, under energy-bounded noise, it attains the best achievable worst-case accuracy on each parameter for a suitable regularization weight. The numerical results demonstrate that the proposed orthogonalization yields consistent physical parameter estimates across different kernels and data sets, while preserving prediction accuracy comparable to that of standard KME. Physical parameters estimated with different kernels or from different experiments thus refer to the same target, fixed by the reference geometry before estimation.

Future work will focus on extending the framework to models that are nonlinear in their parameters and on developing data-driven methods for selecting the discrepancy bound and, through it, the regularization weight that attains the best achievable accuracy.

\section*{Acknowledgements}
This work was supported by the Italian Ministry of University and Research (MUR) under the Fondo Italiano per la Scienza (FIS 2) program, through the project ``ANTSat - Morphing from sparse everywhere-coverage to dense high-capacity: The ANTSat form factor, a deep-space reconfigurable communication system'' [grant number FIS-2023-00667, CUP B53C25003400001].

\bibliographystyle{unsrt}
\bibliography{refs}

\clearpage
\setcounter{section}{0}\setcounter{equation}{0}\setcounter{figure}{0}\setcounter{table}{0}
\renewcommand{\thesection}{S\arabic{section}}
\renewcommand{\theequation}{S\arabic{equation}}
\renewcommand{\thefigure}{S\arabic{figure}}
\renewcommand{\thetable}{S\arabic{table}}
\renewcommand{\theHsection}{S\arabic{section}}
\renewcommand{\theHequation}{S\arabic{equation}}
\fancyhead[R]{\small\itshape Supplementary material}
\thispagestyle{first}
\begin{center}
{\Large\bfseries Supplementary material\par}
\vspace{0.7em}
{\large Separating physics from discrepancy: Orthogonalized kernels for physics-encoded system identification\par}
\vspace{0.7em}
C.~Donati, C.~Novara, F.~Dabbene, M.~Mammarella
\end{center}
\vspace{1.2em}

\section{Computation of the moment representers}\label{app:closedform}
\subsection{Mat\'ern kernels}

Let $\Omega=[a,b]$, $\rho\equiv1$, and $f_j(\xi)=\xi^m$; a polynomial basis follows by linearity. The Mat\'ern kernel of order $q+1/2$ is $\kappa(\xi,x)=P(\alpha|\xi-x|)\,e^{-\alpha|\xi-x|}$, with $\alpha\doteq\sqrt{2q+1}/\ell$ and $P(z)=\sum_{s=0}^{q}p_sz^s$, where $P(z)=1$, $1+z$, and $1+z+z^2/3$ for $q=0,1,2$, respectively. Since the integrand in \eqref{eqn:ok.eta} is not smooth at $\xi=x$, the integral is split at this point. Writing $\xi=x+\sigma r$, with $r\geq0$ and $\sigma=\pm1$ on the two sides, the binomial theorem gives, for every $x\in\mathbb{R}$,
\begin{equation}
\eta_j(x)=\sum_{\sigma=\pm1}\sum_{n=0}^{m}\sum_{s=0}^{q}\binom{m}{n}x^{m-n}\sigma^np_s\alpha^s\,I_{n+s}(r^-_\sigma,r^+_\sigma),
\label{eqn:app.matern}
\end{equation}
where, with $(\cdot)_+\doteq\max\{\cdot,0\}$, $(r^-_{1},r^+_{1})=((a-x)_+,(b-x)_+)$, $(r^-_{-1},r^+_{-1})=((x-b)_+,(x-a)_+)$, and
\[
I_k(r^-,r^+)\doteq\int_{r^-}^{r^+}r^ke^{-\alpha r}dr=\frac{k!}{\alpha^{k+1}}\bigl[Q_k(\alpha r^-)-Q_k(\alpha r^+)\bigr],
\]
with $Q_k(z)\doteq e^{-z}\sum_{l=0}^{k}z^l/l!$, since $Q_k'(z)=-z^ke^{-z}/k!$. The expression \eqref{eqn:app.matern} also holds for $x\notin\Omega$, i.e., for data outside the operating domain. For $x\in\Omega$, $\eta_j$ is a combination of $x^i$, $x^ie^{\alpha x}$, and $x^ie^{-\alpha x}$; hence, $H=\int_\Omega f(x)\eta(x)^\top dx$ is obtained from the same integrals or, to roundoff, by a Gauss--Legendre rule.

\subsection{Several inputs and other kernels}

Let $\Omega=\Omega_1\times\dots\times\Omega_{n_x}$ be a box, $\rho(\xi)=\prod_d\rho_d(\xi_d)$, $f_j(\xi)=\prod_df_{jd}(\xi_d)$, and $\kappa(\xi,x)=\prod_{d\in\mathcal{I}}\kappa_d(\xi_d,x_d)$, with $\mathcal{I}\subseteq\{1,\dots,n_x\}$. Then, with $\bar f_{jd}\doteq\int_{\Omega_d}f_{jd}\rho_d\,d\xi_d$,
\[
\eta_j(x)=\prod_{d\in\mathcal{I}}\int_{\Omega_d}f_{jd}(\xi_d)\kappa_d(\xi_d,x_d)\rho_d(\xi_d)\,d\xi_d\prod_{d\notin\mathcal{I}}\bar f_{jd},
\]
and $H_{ij}$ factorizes in the same way, the coordinates $d\notin\mathcal{I}$ contributing $\bar f_{id}\bar f_{jd}$. For monomial $f_{jd}$ and $\rho_d\equiv1$, the factors with $d\in\mathcal{I}$ are those of the previous subsection. If $\bar f_{jd}=0$ for some $d\notin\mathcal{I}$, then $\eta_j\equiv0$ and $H$ is singular, as in Section~\ref{sec:num.msd}.

For a polynomial kernel $\kappa(\xi,x)=\sum_kc_k\xi^kx^k$, $c_k\geq0$, the variables separate: $\eta_j(x)=\sum_kc_k\langle f_j,\xi^k\rangle_\rho x^k$ and $H_{ij}=\sum_kc_k\langle f_i,\xi^k\rangle_\rho\langle f_j,\xi^k\rangle_\rho$. For the Gaussian kernel $\kappa(\xi,x)=e^{-(\xi-x)^2/(2\ell^2)}$, with $\Omega$, $\rho$, and $f_j$ as above, no splitting is needed and $\eta_j(x)=\sum_{n=0}^{m}\binom{m}{n}x^{m-n}M_n(x)$, where $M_n(x)\doteq\int_{a-x}^{b-x}u^ne^{-u^2/(2\ell^2)}\,du$. Here, $M_0$ is given by the error function, and, for $n\geq1$, $M_n=(n-1)\ell^2M_{n-2}-\ell^2\bigl[u^{n-1}e^{-u^2/(2\ell^2)}\bigr]_{a-x}^{b-x}$, the first term being absent for $n=1$; integration by parts yields $H$ in terms of the same functions.

\subsection{Quadrature}

The rule in \eqref{eqn:impl.quad} is typically composite Gauss--Legendre, with panels split at the discontinuities of $f$ and $\rho$, assumed bounded and piecewise continuous. As the panel width vanishes, $\eta_M\to\eta$ uniformly on compact sets and $H_M\to H$; hence, if $H\succ0$, the kernel obtained from \eqref{eqn:impl.quad} converges to $\kappa_\perp$. The condition $H\succ0$ holds if the restriction of $\mathcal{H}$ to $\Omega$ is dense in $C(\Omega)$: by \eqref{eqn:ok.riesz}, $c^\top Hc=0$ implies $\langle f^\top c,\delta\rangle_\rho=0$ for every $\delta\in C(\Omega)$; hence, since $C(\Omega)$ is dense in $L^2_\rho$, $c^\top Gc=\|f^\top c\|_\rho^2=0$, and $c=0$ by Assumption~\ref{ass:ok.geometry}. This density holds for Gaussian and Mat\'ern kernels, and for their products over all the coordinates of $x$, since their Fourier transforms are positive~\cite[Proposition~15]{micchelli2006universal}. If $H$ is singular, $H_M^\dagger$ need not converge to $H^\dagger$, since the pseudoinverse is discontinuous where the rank changes; a truncated pseudoinverse, with threshold below the smallest positive eigenvalue of $H$, converges.

With fixed nodes, the convergence for Mat\'ern kernels is only algebraic, since, for $x\in\Omega$, the integrand in \eqref{eqn:ok.eta} has a discontinuous derivative of order $2q+1$ at $\xi=x$, in general inside a panel. Therefore, in practice, for product kernels, the panels are also split at $\xi_d=x_d$ in each coordinate, so that, for $f$ and $\rho$ analytic on each panel, the error decreases geometrically with the number of nodes~\cite[Theorem~4.5]{trefethen2008gauss}, and $H$ is computed from $\eta$ by a fixed rule with the same discontinuities. The nodes then depend on $x$, and the discrete inner product of Section~\ref{sec:impl.quad} no longer applies: the constraints in \eqref{eqn:ok.Hperp} and the positive semidefiniteness of $\kappa_\perp$ hold up to the quadrature error, which, for the constraints, is measured by the verification of Section~\ref{sec:impl.quad}. This is the rule used in Section~\ref{sec:num} when no closed form is available. In any case, the nodes must not be the data: if they coincide with the inputs $x_t$, with equal weights, then $F(x)^\top\mathbf{K}_\perp=0$, and $\theta^\star_\perp$ reduces to the least-squares estimate for every kernel.

\end{document}